\documentclass[fleqn,usenatbib]{mnras}
 
\usepackage{graphicx} 
\usepackage{txfonts} 
\usepackage{natbib}               
\title[] {A chromosome map to unveil stellar populations with different magnesium abundances. The case of $\omega$\,Centauri.}
\author[ A.\,P.\,Milone et al.] 
       {A.\,P.\,Milone$^{1,2}$, A.\,F.\,Marino$^{1,2,3}$, A.\,Renzini$^2$, C.\,Li$^{4}$, 
       S.\,Jang$^5$, E.\,P.\,Lagioia$^{1}$,\newauthor
       M.\,Tailo$^{1}$, G.\,Cordoni$^{1}$,   M.\, Carlos$^{6}$, E.\,Dondoglio$^{1}$ \\ 
$^{1}$Dipartimento di Fisica e Astronomia ``Galileo Galilei'', Univ. di Padova, Vicolo dell'Osservatorio 3, Padova, IT-35122\\
$^{2}$Istituto Nazionale di Astrofisica - Osservatorio Astronomico di Padova, Vicolo dell'Osservatorio 5, Padova, IT-35122\\
$^{3}$ Centro di Ateneo di Studi e Attivita' Spaziali ``Giuseppe Colombo'' - CISAS, Via Venezia 15, Padova, IT-35131 \\
$^{4}$ School of Physics and Astronomy, Sun Yat-sen University, Zhuhai 519082, People's Republic of China.\\
$^{5}$ Center for Galaxy Evolution Research and Department of Astronomy, Yonsei University, Seoul 03722, Korea\\
$^{6}$ Departamento de Astronomia, IAG, Universidade de S\~{a}o Paulo, Rua do Mat\~{a}o 1226, 05509-900 S\~{a}o Paulo, Brazil
} 
\begin{document} 
 
\maketitle 
\label{firstpage}

\begin{abstract}
 Historically, photometry has been largely used to identify stellar populations (MPs) in Globular Clusters (GCs) by using diagrams that are based on colours and magnitudes that are mostly sensitive to stars with different metallicities or  different abundances of helium, carbon, nitrogen, and oxygen. In particular, the pseudo two-colour diagram called chromosome map (ChM), allowed the identification and the characterization of MPs in about 70 GCs by using appropriate filters of the {\it Hubble Space Telescope} ({\it HST}) that are sensitive to the stellar content of He, C, N, O and Fe. 
 
 We use here high precision {\it HST} photometry from F275W, F280N, F343N, F373N, and F814W images of $\omega$\,Centauri to investigate its MPs.
 We introduce a new ChM whose abscissa and ordinate are mostly sensitive to stellar populations with different magnesium and nitrogen, respectively, in monometallic GCs. 
 This ChM is effective in disentangling the MPs based on their Mg chemical abundances, allowing us to explore, for the first time, possible relations between the production of these elemental species for large samples of stars in GCs.

 By comparing the colours of the distinct stellar populations with the colours obtained from appropriate synthetic spectra we provide "photometric-like" estimates of the chemical composition of each population.
 Our results show that, in addition to first generation (1G) stars, the metal-poor population of $\omega$\,Cen hosts four groups of second-generation stars with different [N/Fe],  namely, 2G$_{\rm A--D}$. 2G$_{\rm A}$ stars share nearly the same [Mg/Fe] as the 1G, whereas 2G$_{\rm B}$, 2G$_{\rm C}$ and 2G$_{\rm D}$ are Mg depleted by $\sim$0.15, $\sim$0.25 and $\sim$0.45 dex, respectively. We provide evidence that the metal-intermediate populations host stars with depleted [Mg/Fe].

\end{abstract} 
 

\section{Introduction}\label{sec:intro}
 The color-magnitude diagrams (CMDs) of most Galactic and extragalactic globular clusters (GCs) are composed of distinct photometric sequences that  are detected among stars at different evolutionary stages and correspond to stellar populations with different content of helium and light elements \citep[][and references therein]{milone2012a, milone2020a}. Specifically, most GCs host a first generation (1G) of stars that share the same chemical composition as field stars with similar metallicities and second stellar generations (2G) that are enhanced in helium nitrogen and sodium and depleted in carbon and oxygen \citep[e.g.][and references therein]{kraft1993a, gratton2012a, bastian2018a, milone2018a, marino2019a}. 

The star-to-star variation of light elements and the corresponding correlations and anticorrelations observed in GCs   \citep[e.g.\,][]{hesser1977a, kraft1993a, Ivans2001a, marino2008a, carretta2009a, meszaros2015a}  are understood as the results of CNO cycling and p-capture processes at high temperatures. These processes include the Mg-Al chain, effective at temperatures higher than $\sim 7 \times 10^{7}$ K, which produces Al at the expenses of Mg and is responsible for the Mg-Al anticorrelations observed in some GCs  from spectroscopy \citep[e.g.\,][]{carretta2015a, masseron2019a}. 
Indeed, Mg abundances are proxies of stellar nucleosynthesis processes occurring at higher temperature than those responsible for the classic, and much more widespread, N-C/Na-O anticorrelations. 

 In addition to internal variations of light elements, some GCs exhibit stellar populations with different content of heavy elements, including iron and s-process elements \citep[e.g.][]{marino2009a, marino2015a, dacosta2009a, carretta2010a, yong2014a, johnson2017a}.  

 Clusters with heavy element variations are called Type~II GCs and comprise $\sim$17\% of the total number of Galactic GCs \citep{milone2017a, milone2020a}.  A common feature of Type-II GCs is that stars with similar metallicities host stellar populations with different light-elements abundances   \citep[e.g.][]{marino2009a, marino2011a, milone2015a, milone2017a}.

 The origin of MPs is one of the most-debated open issues of stellar astrophysics, which could have significantly affected the assembly of the Galactic halo and possibly the reionization of the Universe \citep[e.g.][]{renzini2015a, renzini2017a}. \\
 According to some scenarios, GCs have experienced  multiple   star-formation episodes and 2G stars formed  out of the material  ejected by  more-massive 1G stars \citep[e.g.][]{ventura2001a, decressin2007a, dercole2008a, denissenkov2014a, dantona2016a, calura2019a}. 
Alternatively, GC stars are coeval and  exotic phenomena that occurred in the unique environment of the proto GCs are responsible for the chemical composition of 2G stars  \citep[e.g.][]{bastian2013a, gieles2018a}. We refer to \citet{renzini2015a} for critical discussion on the various scenarios.

Various photometric diagrams have been exploited in the past decade to disentangle multiple populations (MPs) in Globular Clusters (GCs) and allowed to identify and characterize MPs in more than seventy Galactic and extragalactic clusters \citep[e.g.][]{milone2017a}. Most of these diagrams are based on $U$-band photometry, which allows to distinguish 1G from 2G stars mainly through the effect of NH molecules on the ultraviolet stellar flux \citep[e.g.][]{marino2008a, yong2008a, milone2012b, milone2015a}.    

Some work is based on CMDs made with a wide colour baseline, which is sensitive to stellar populations with different helium abundances. Indeed, main sequence (MS) and red giant branch (RGB) stars enhanced in helium  are hotter, hence bluer than stars with the same luminosity and primordial helium (Y$\sim0.25$) \citep[e.g.][]{dantona2002a, bedin2004a, piotto2007a, milone2018a}. Similarly, the $U-V$ or $U-I$ colours are widely used to distinguish stellar populations with different total metal content \citep[Z, e.g.][]{marino2015a, milone2017a}.

\citet{milone2015a} introduced a new photometric diagram called chromosome map (ChM) to identify and characterize the distinct stellar populations of GCs.  The ChM is a pseudo two-colour diagram of MS, RGB or asymptotic-giant branch (AGB) stars, where the photometric sequences are verticalized in both dimensions. It maximizes the separation between stellar populations with different He, C, N, O and Fe.

Near infrared photometry is a powerful tool to identify MPs of M-dwarfs with different oxygen abundances and the F110W and F160W filters of the WFC3/NIR camera on board of {\it HST} are the most widely used filters \citep[e.g.\,][]{milone2012a, milone2019a}. Indeed, the F160W band is heavily affected by absorption from various molecules involving oxygen, including H$_{2}$O, while F110W photometry is almost unaffected by the oxygen abundance. As a consequence, second-generation (2G) stars, which are depleted in O with respect to the 1G, have brighter F160W magnitudes and redder F110W$-$F160W colours than the 1G.

In summary, all photometric diagrams used to detect multiple populations are based on colours and magnitudes that are mostly sensitive to the abundances of helium, nitrogen, oxygen and metallicity of stars in the distinct populations.  

While photometry has been successful in identifying stellar populations with different C/N/O/Na in GCs, to date, it was almost blind to stars that are composed of material exposed to higher-temperature H burning, e.g. depleted in Mg with respect to the 1G.
In this paper, we introduce new photometric diagrams to disentangle stars based on their  magnesium abundances. This is the first time that multiple populations with different [Mg/Fe] are identified from photometry alone.

 Our target, $\omega$\,Centauri is the GC where multiple populations have been first detected \citep{woolley1966a} 
 and one of the most-studied clusters in the context of MPs  \citep[e.g.][]{anderson1997a,lee1999a, pancino2000a, bedin2004a, ferraro2004a, sollima2005a, sollima2007a, bellini2010a}. 
 
  $\omega$~Centauri  is an extreme Type II GC, which hosts at least sixteen populations that span a wide interval of metallicity,  
  (from [Fe/H] $\lesssim -2.0$ to [Fe/H]$\sim -0.6$) 
  \citep[e.g.][]{freeman1975a,  suntzeff1996a, norris1995a, norris1996a, johnson2010a, marino2011a}, reach extreme abundances of helium and light elements \citep[e.g.][]{norris2004a, tailo2016a, bellini2017a, milone2017b}, and exhibit star-to-star magnesium variations \citep{gratton1982a, norris1995a, dacosta2013a, meszaros2019a}. 
   In particular, similarly to what is observed in other Type-II GCs,  stellar populations of  $\omega$~Centauri with  any given metallicity span a wide range in light-elements abundances \\ \citep[e.g.][]{marino2010a, marino2011b, marino2012a, johnson2010a, gratton2011a}.


The paper is organized as follows. Section~\ref{sec:data} describes the dataset and the data reduction to derive the photometric diagrams presented in Section~\ref{sec:dia}. The theoretical ChMs and the photometric diagrams from isochrones are discussed in Section~\ref{sec:teo}, while Section~\ref{sec:mpoor} and Section~\ref{sec:mrich} are focused on the metal-poor and metal-rich stellar populations of $\omega$\,Cen, respectively. Summary and conclusions are provided in Section~\ref{sec:summary}.

\section{Data and data analysis}\label{sec:data}
The main dataset used in this work is composed of images collected through the F275W, F280N, F343N, F373N and F814W filters of the Ultraviolet and Visual Channel of the Wide Field Camera 3 (UVIS/WFC3) on board of {\it Hubble Space Telescope} ({\it HST}). The main properties of these images are summarized in Table~\ref{tab:data}. 
Photometry and astrometry have been performed from images corrected from the poor charge transfer efficiency \citep[see][]{anderson2010a}. We used the computer program KS2, 
that is the evolution of $kitchen\_sink$, originally written to reduce two-filter images collected with the Wide-Field Channel of the Advanced Camera for Survey of {\it HST} \citep{anderson2008a}.
As discussed by \citet{sabbi2016a} and \citet{bellini2017a}, KS2 adopts two different methods to derive high-precision measurements of stars with different luminosities. To determine magnitudes and positions of faint stars we combine information from all exposures and determine the average stellar positions. Once the positions of faint stars are fixed we fit each exposure pixel with the appropriate effective point spread function (PSF) solving for the flux only.
Bright stars are measured in each individual exposure by fitting the best PSF model and the resulting magnitudes and positions are then averaged.

Instrumental magnitudes are calibrated into the Vega-mag system by using the updated photometric zero points\footnote{\url{http://www.stsci.edu/hst/wfc3/analysis/uvis_zpts/} and \url{http://www.stsci.edu/hst/acs/analysis/zeropoints}} and following the recipe by \citet{bedin2005a}. Stellar positions are corrected for geometrical distortion by using the solutions provided by \citet{bellini2009a} and \citet{bellini2011a}. Finally, photometry was corrected for differential reddening as in \citet{milone2012a}.

To increase the number of filters and better constrain the abundances of helium, carbon, nitrogen, oxygen, and magnesium of the stellar populations of $\omega$\,Cen, we used the photometric and astrometric catalogues from \citet{milone2017a} and \citet{milone2018a}, which include photometry obtained from images collected through 31 filters of UVIS/WFC3.
 Details of the dataset and the data reduction are provided by \citet{milone2017a} and \citet{milone2018a}.
In summary, our database comprises photometry in 36 bands: F218W, F225W, F275W, F280N, F300X, F336W, F343N, F373N, F390M, F373N, F390M, F390W, F395N, F410M, F438W, F467M, F469N, F475W, F487N, F502N, F555W, F606W, F621M, F625W, F631M, F645M, F656N, F657N, F658N, F673N, F673N, F680N, F689M, F763M, F680N, F689M, F763M, F775W, F814W, F845M and F953N.

\begin{table*}
  \caption{Description of the {\it HST} images used in the paper.}

\begin{tabular}{ c c c l l}
\hline \hline
 FILTER  & DATE & N$\times$EXPTIME & PROGRAM & PI \\
\hline
  F275W & Jul 15 2009  & 35$+$9$\times$350s       & 11452 & J.\,Kim\,Quijano\\
  F275W & Jan 12 -- Jul 4 2010   & 22$\times$800s & 11911 &  E.\,Sabbi   \\ 
  F275W & Feb 02 2011  &  9$\times$800s & 12339 &   E.\,Sabbi  \\  
  F280N & Feb 26 2015  &  600s & 14031 &  V.\,Kozhurina-Platais  \\  
  F280N & Feb 01 2016  &  2$\times$800$+$4$\times$850s & 14393 &  V.\,Kozhurina-Platais  \\  
  F343N & Feb 26 2015  &  600s & 14031 &  V.\,Kozhurina-Platais  \\  
  F343N & Mar 03 2016  &  5$\times$510$+$4$\times$545s & 14393 &  V.\,Kozhurina-Platais \\  
  F373N & Feb 26 2015  &  450s & 14031 &  V.\,Kozhurina-Platais  \\  
  F373N & Mar 25 2016  &  5$\times$500s & 14393 &  V.\,Kozhurina-Platais \\  
  F814W & Jul 15 2009 & 35s & 11452 & J.\,Kim Quijano\\
  F814W & Jan 10 -- Jul 04 2010 & 27$\times$40s & 11911 & E.\,Sabbi\\
  F814W & Feb 15 -- Mar 24 2011 & 9$\times$40s &  12339 & E.\,Sabbi\\

     \hline\hline
\end{tabular}
  \label{tab:data}
 \end{table*}

\begin{centering} 
\begin{figure} 
  \includegraphics[height=10.cm,trim={0.0cm 5cm 5cm 3.0cm},clip]{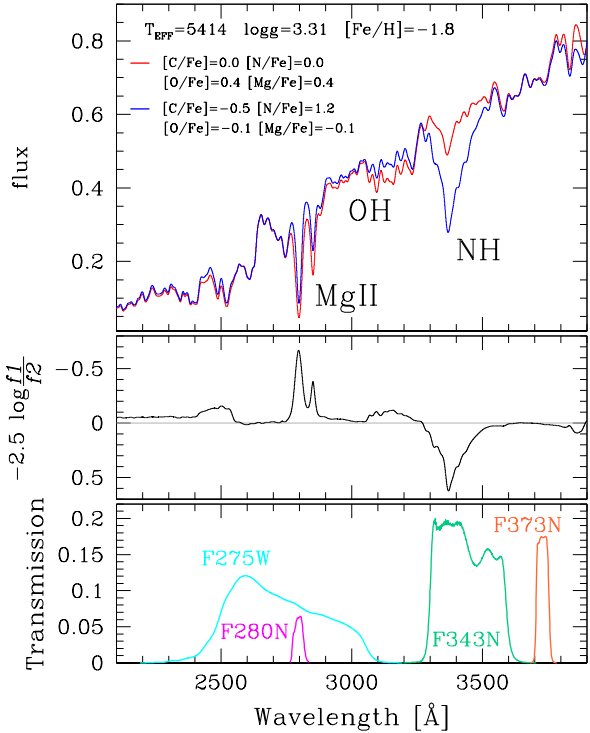}
 \caption{Comparison of two synthetic spectra with the same stellar parameters and metallicity quoted in the inset but different abundances of light elements (top panel). The main spectral features that are responsible for the differences between the fluxes of the two spectra are also indicated. In the middle panel we plot the quantity $-2.5 \log{f1/f2}$ as a function of the wavelength, where $f1$ and $f2$ are the fluxes of the red and the blue spectrum, respectively. Lower panel shows the transmission curves of the F275W, F280N, F343N and F373N WFC3/UVIS filters.}
 \label{fig:spettri} 
\end{figure} 
\end{centering} 
\section{Photometric diagrams}\label{sec:dia}

In this section we construct the photometric diagrams that will be exploited to analyse stellar populations with different Mg in $\omega$~Cen, starting from of the spectral features most sensitive to this element. 
Based on synthetic spectra of RGB stars with different chemical composition, \citet{milone2018a} show that [Mg/Fe] variations have a negligible effect on magnitudes in optical bands but provide significant  flux variations in ultraviolet bands  \citep[see also][]{sbordone2011a}. 

As an example, in the upper panel of Figure~\ref{fig:spettri} we compare the spectra of two RGB stars that have the same chemical abundances in all the elements, except for C, N, O and Mg, as quoted in the figure.  
Specifically, the red spectrum is indicative of a 1G stars, whereas the blue spectrum corresponds to the 2G.
We calculated the ratio between the fluxes ($f1$ and $f2$) of the blue and red spectra and plotted in the middle panel of Figure~\ref{fig:spettri} the quantity $-2.5 \log{f1/f2}$ against $\lambda$. Lower panels show the transmission curves of the F275W, F280N, F343N and F373N filters of WFC3/UVIS. 

Clearly, the F280N band, which includes the strong Mg-II lines at $\lambda$ $\sim$ 2,800 \AA\ is the most efficient filter on board of {\it HST} to identify stars with different values of [Mg/Fe]. 
The Mg-rich star is about 0.25 mag fainter in F280N than the Mg-poor one, the flux difference drops to $\sim$0.02 in F275W. Hence, $m_{\rm F275W}-m_{\rm F280N}$ is an efficient colour to identify stellar populations with different magnesium abundances.

Moreover, Figure~\ref{fig:spettri} reveals that the $m_{\rm F343N}-m_{\rm F373N}$  colour is sensitive to stellar populations with different nitrogen abundances. Indeed, the F343N filter includes the NH molecular bands around $\lambda \sim 3370$\AA, whereas the spectral region covered by the F373N filter is poorly affected by light-element variations. 

Since $m_{\rm F275W}-m_{\rm F280N}$ and $m_{\rm F343N}-m_{\rm F373N}$ are short colour baselines, they are both poorly sensitive to  temperature and reddening differences among stars with similar luminosities.

\begin{centering} 
\begin{figure*} 
  \includegraphics[height=6.cm,trim={0.5cm 6.1cm 0cm 9.2cm},clip]{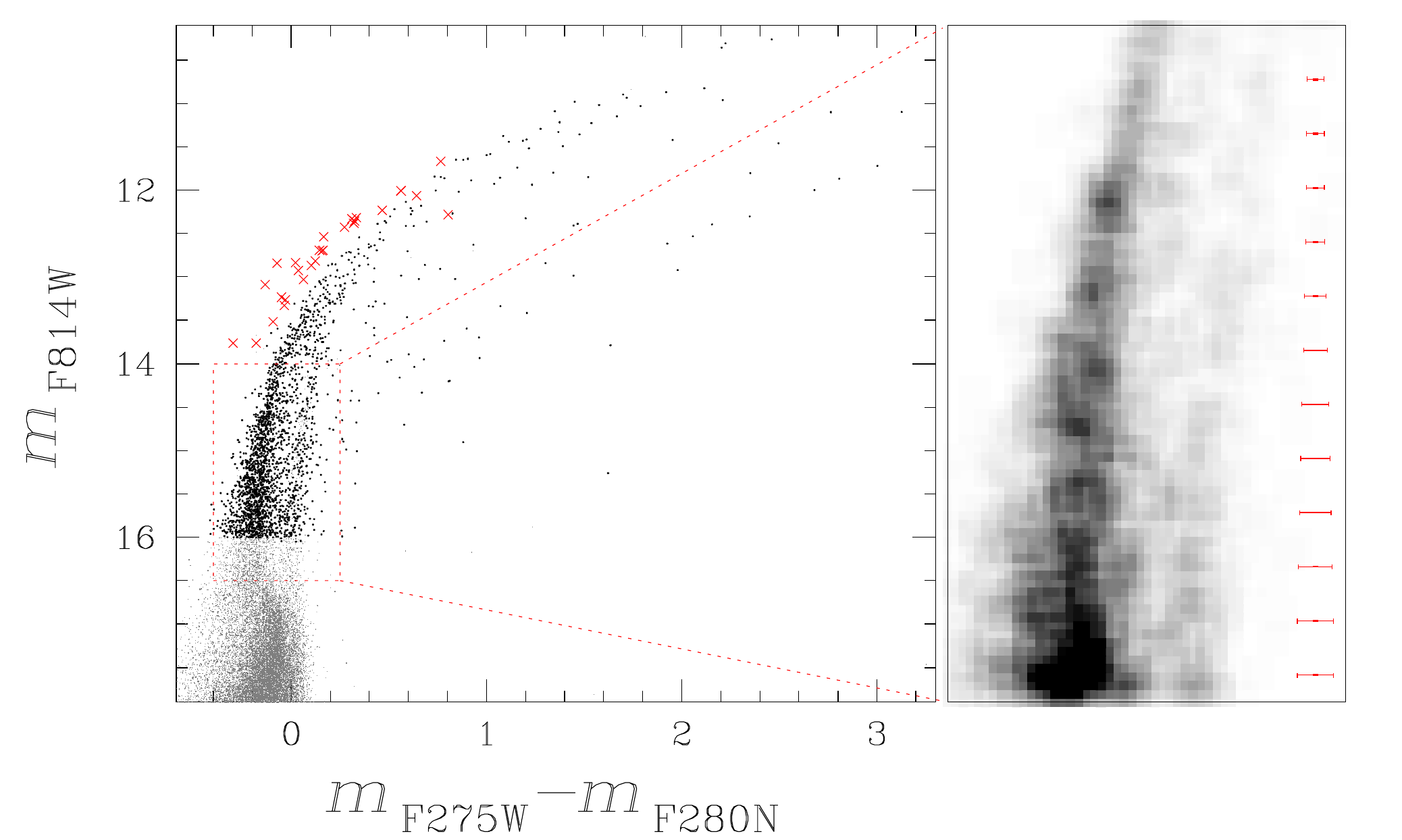}
  \includegraphics[height=6.cm,trim={0.5cm 6.1cm 0cm 9.2cm},clip]{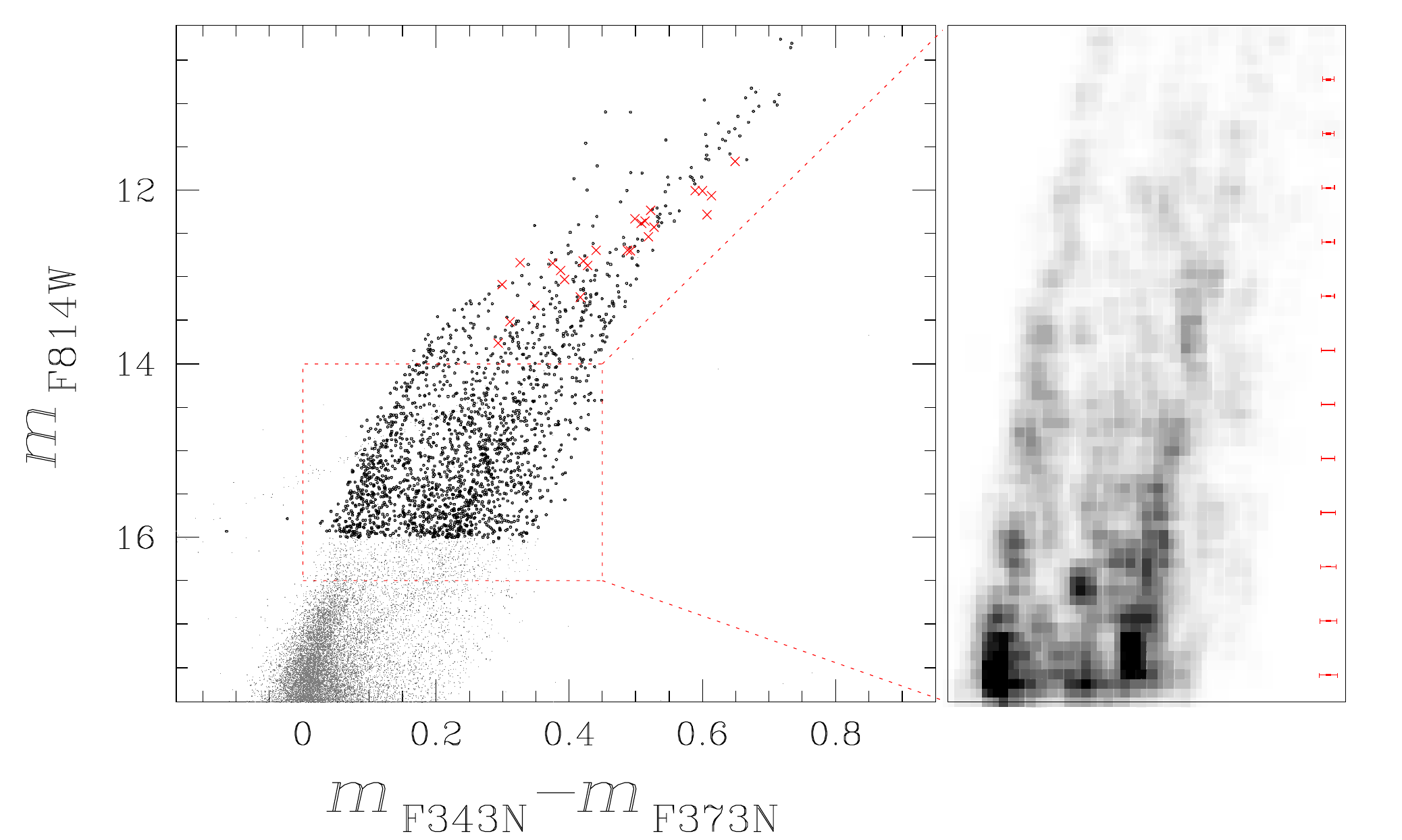}
  \caption{Comparison between the $m_{\rm F814W}$ vs.\,$m_{\rm F275W}-m_{\rm F280N}$ CMD of of $\omega$\,Cen (top), which is very sensitive to stellar populations with different magnesium content, and the $m_{\rm F814W}$ vs.\,$m_{\rm F343N}-m_{\rm F373N}$ CMD (bottom), which highlights stars with different nitrogen abundance. RGB stars brighter than $m_{\rm F814W}=16.5$ are marked with black dots, while red crosses indicate AGB stars. Right panels show the Hess diagrams for stars in the rectangular regions shown in the right panels. }
 \label{fig:cmds} 
\end{figure*} 
\end{centering}

The observed $m_{\rm F814W}$ vs.\,$m_{\rm F275W}-m_{\rm F280N}$ CMD of $\omega$\,Cen is shown in the upper-left panel of Figure~\ref{fig:cmds}. The RGB exhibits two main parallel sequences in the F814W magnitude interval between $\sim$16.5 and $\sim$13.8, as highlighted in the Hess-diagram plotted in the upper-right panel. The two RGB components seem to merge together at $m_{\rm F814W} \lesssim 13.8$. Additional RGB stars exhibit redder $m_{\rm F275W}-m_{\rm F280N}$ colours than the bulk of $\omega$\,Cen RGB stars. Their colour distance from the main RGB increases when we move from the base of the RGB towards the RGB tip.     

The lower panels of Figure~\ref{fig:cmds} show the $m_{\rm F814W}$ vs.\,$m_{\rm F343N}-m_{\rm F373N}$ CMD  for $\omega$\,Cen RGB stars (left) and the corresponding Hess diagram for RGB stars with $14.0<m_{\rm F814W}<16.5$.  
RGB stars of $\omega$\,Cen span a wide range of $m_{\rm F343N}-m_{\rm F373N}$ and exhibit four main stellar overdensities that indicate stellar populations with different nitrogen content.

For completeness, we mark with red crosses the asymptotic giant branch stars selected from the $m_{\rm F814W}$ vs.\,$m_{\rm F606W}-m_{\rm F814W}$ CMD. Since AGB stars have comparable stellar parameters as RGB stars,
a variation in light elements would produce similar broadening 
in the AGB sequences. The fact that the AGB is narrower than the RGB in both CMDs of Figure~\ref{fig:cmds} 
suggests
that stellar populations with extreme contents of nitrogen and magnesium avoid the AGB phase, in close analogy with what is observed in other GCs \citep[e.g. NGC\,6752, NGC\,2808 and NGC\,6266][]{campbell2013a, wang2016a, marino2017a, lapenna2015a}. 

\section{Comparison with simulated multiple populations}\label{sec:teo}

As discussed in the Introduction, $\omega$\,Cen exhibits large star-to-star metallicity variations, and the stellar populations with different [Fe/H] host sub stellar populations with distinct content of helium and light element abundance.   
To investigate the physical reasons that are responsible for the multiple RGBs shown in Figure~\ref{fig:cmds}, we first compare in Section~\ref{sub:FeTeo} isochrones from the Dartmouth database \citep{dotter2008a} with different metallicities and helium contents.
Then, in Section~\ref{sub:NTeo} we investigate the effect of light-element abundance variations on the CMDs by using mono-metallic isochrones with different content of He, C, N, O and Mg.

\subsection{Stellar populations with different metallicities}\label{sub:FeTeo}
\begin{centering} 
\begin{figure*} 
  \includegraphics[height=9.cm,trim={.7cm 5cm 1cm 5cm},clip]{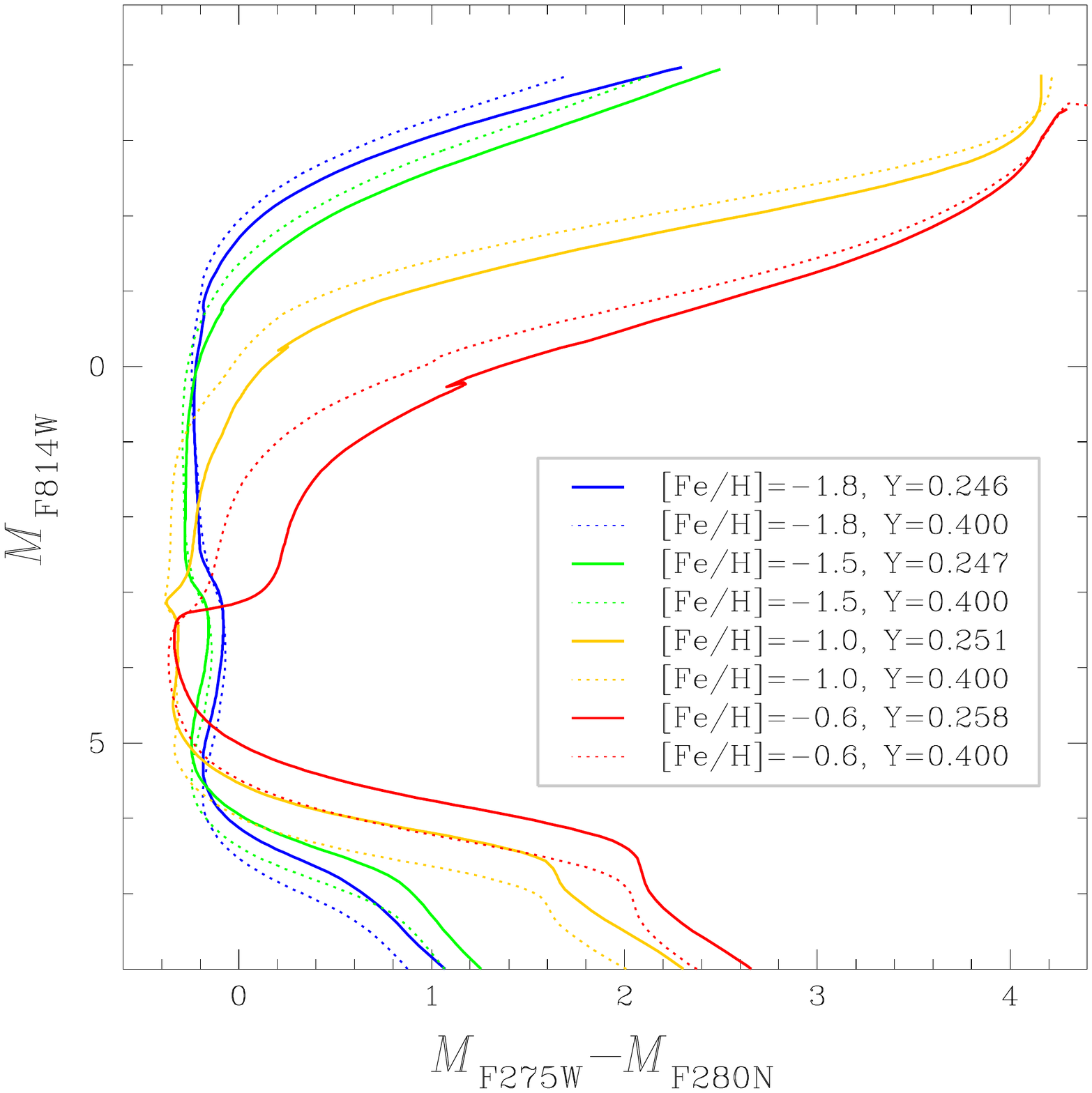}
  \includegraphics[height=9.cm,trim={.7cm 5cm 1cm 5cm},clip]{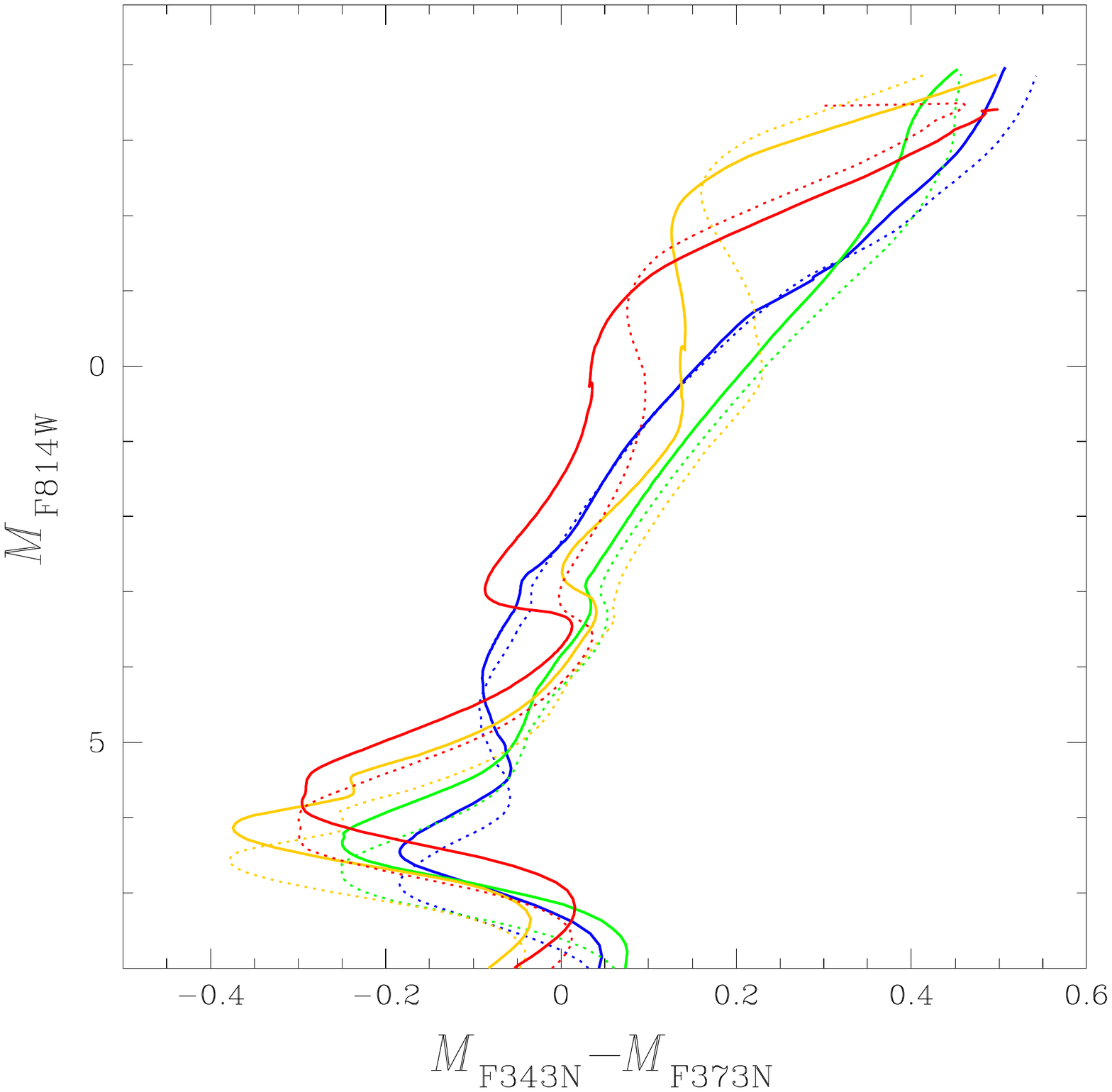}
  \caption{Dartmouth isochrones \citep{dotter2008a} in the  $M_{\rm F814W}$ vs.\,$M_{\rm F275W}-M_{\rm F280N}$  (left panel) and $M_{\rm F814W}$ vs.\,$M_{\rm F343N}-M_{\rm F373N}$ (right panel) planes for 13-Gyr old stellar populations with [$\alpha$/Fe]=0.4  and with the abundances of iron and helium listed in the inset of left panel. The RGB sequences comprise the isochrone segments brighter than $M_{\rm F814W} \sim 3.0$. }
 \label{fig:iso} 
\end{figure*} 
\end{centering}

Figure~\ref{fig:iso} shows the $M_{\rm F814W}$ vs.\,$M_{\rm F275W}-M_{\rm F280N}$ (left panel) and the $M_{\rm F814W}$ vs.\,$M_{\rm F343N}-M_{\rm F373N}$ (right panel) CMDs for isochrones from \citet{dotter2008a} with different iron and helium abundances.
Specifically, we considered isochrones with [Fe/H]=$-1.8$ and [Fe/H]=$-1.5$, which correspond to the two main peaks in the metallicity distribution based on high-resolution spectroscopy \citep{marino2011a}, that are represented with blue and green colours, respectively. Moreover, we used isochrones with [Fe/H]=$-1.0$ and [Fe/H]=$-0.6$, which are representative to the most metal rich stellar populations of $\omega$\,Cen.  
The isochrones represented with continuous lines have helium abundance Y=0.25+1.5 Z, where Z is the total metal abundance, while those isochrones plotted with dotted lines have helium Y=0.40.

 A significant separation along the RGB is apparent at very high-metallicity and in the upper portion of the RGB with the metal-rich isochrones characterised by redder $M_{\rm F275W}-M_{\rm F280N}$ colours than metal-poor ones.
The magnitude where the metal-rich isochrones cross the metal poor ones depends on the value of [Fe/H]. As an example, the isochrones with [Fe/H]=$-1.5$, [Fe/H]=$-1.0$ and [Fe/H]=$-0.6$  intersect the  [Fe/H]=$-1.8$ isochrone at $M_{\rm F814W} \sim 0.0$, $\sim 0.5$ and $\sim$2.7, respectively. An exception is provided by the isochrone with [Fe/H]=$-0.6$, which is redder than the used metal-poor isochrones along the entire RGB.
Helium-rich isochrones have almost the same  $M_{\rm F275W}-M_{\rm F280N}$ colour as the metal-poor ones and become redder at brighter luminosities.

The right panel of Figure~\ref{fig:iso} shows that the low RGBs of the analyzed metal-rich isochrones typically have redder $M_{\rm F343N}-M_{\rm F373N}$ colours than those of metal poor ones. The most metal rich isochrone, which is bluer than isochrones with [Fe/H]=$-1.5$ and [Fe/H]=$-1.0$, is a remarkable exception.   
 The isochrones with different [Fe/H] cross each other and change their relative $M_{\rm F343N}-M_{\rm F373N}$ colours in the upper part of the RGB.
 Helium-rich RGB stars of the isochrones with [Fe/H]$ \geq -1.5$ are redder than stars with Y$\sim$0.25 and the same F814W magnitude, while the RGBs of the two isochrones with [Fe/H]=$-1.8$ and different helium abundances are nearly coincident for $M_{\rm F814W} \gtrsim -1.5$.

\subsection{Stellar populations with different light-element abundances }\label{sub:NTeo}
To investigate how the abundances of C, N, O, Mg affect the position of stars in the CMDs of mono-metallic GCs, we derived the colours and magnitudes of isochrones with different abundances of carbon, nitrogen, oxygen and magnesium by using model atmospheres and synthetic spectra, in close analogy with what done in previous papers from our team \citep[e.g.][]{milone2012b, milone2018a}. 
In a nutshell, we extracted fifteen points along the isochrones and extracted their effective temperatures, $T_{\rm eff}$, and gravities, $\log{g}$. For each pair of stellar parameters we calculated a reference spectrum with the chemical composition of 1G stars and a comparison spectrum, with the content of C, N, O and Mg of 2G stars. 
Specifically, we assumed for 1G stars Y=0.246, solar-scaled abundances of carbon and nitrogen, [O/Fe]=0.4, and [Mg/Fe]=0.4.

 We constructed model atmosphere structures by using the    computer program ATLAS12 developed by Robert Kurucz \citep{kurucz1970a, kurucz1993a, sbordone2004a}, which exploits the opacity-sampling technique and assumes local thermodynamic equilibrium. We assumed a microturbolent velocity of 2 km s$^{-1}$ for all models.
Synthetic spectra are computed over the wavelength interval between 1,000 and 12,000 \AA\ with SYNTHE \citep{kurucz1981a, kurucz2005a, castelli2005a, sbordone2007a} and have been integrated over the bandpasses of the ACS/WFC and UVIS/WFC3 filters used in this paper to derive the corresponding  magnitudes. 
 We calculated the magnitude difference, $\delta {m}_{\rm X}$, between the comparison and the reference spectrum. The magnitudes of the 2G isochrones are derived by adding to the 1G isochrones the corresponding values of $\delta {m}_{\rm X}$. 

\begin{centering} 
\begin{figure*} 
  \includegraphics[height=9.cm,trim={1.0cm 5cm .5cm 12.0cm},clip]{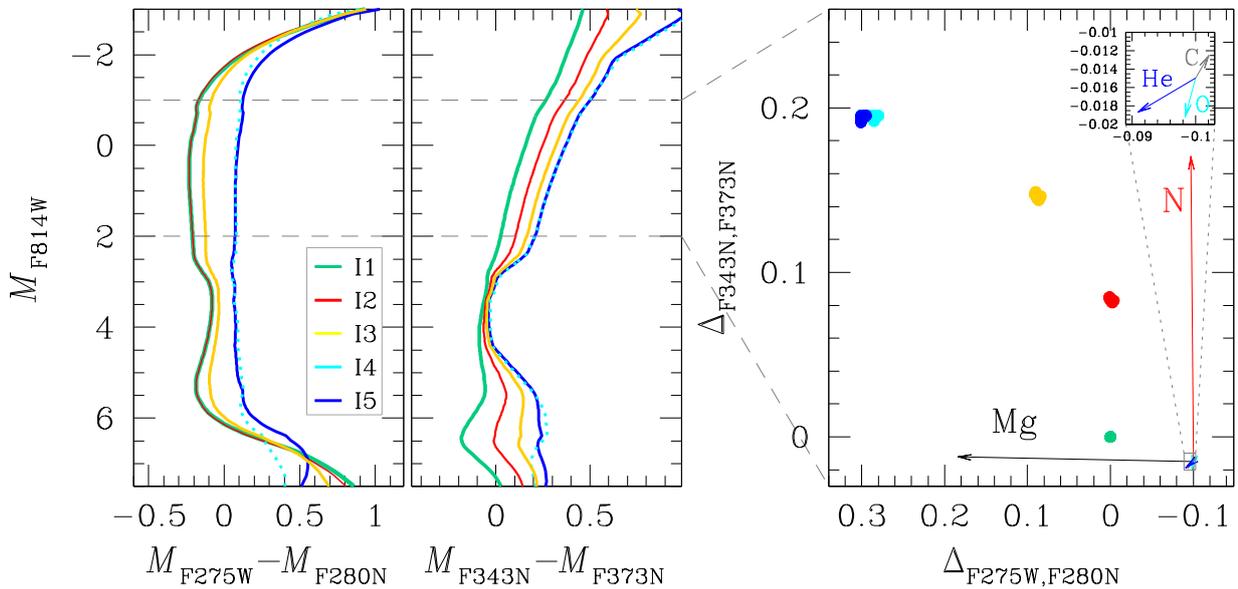}
 \caption{Green, red, yellow, cyan and blue colours represent the isochrones I1--I5, which have ages of 13 Gyr, [Fe/H]=$-1.8$, [$\alpha$/Fe]=0.4 and different abundances of He, C, N, O and Mg as listed in Table~\ref{tab:chimica}.  Right panel show the $\Delta_{\rm F343N,F373N}$ vs.\,$\Delta_{\rm F275W,F280N}$ ChM for RGB stars between the horizontal lines. The arrows indicate the effect of changing Y, C, N, O and Mg one at a time in the ChM. Specifically, we assumed $\Delta$\,Y=0.154, $\Delta$[C/Fe]=$-$0.50, $\Delta$[N/Fe]=1.21, $\Delta$[O/Fe]=$-$0.50, $\Delta$[Mg/Fe]=$-$0.50.  }
 \label{fig:ChM280teo} 
\end{figure*} 
\end{centering}

We plot in Figure~\ref{fig:ChM280teo} five isochrones with the same iron abundance, [Fe/H]=$-1.7$, but different content of He, C, N, O and Mg in the $M_{\rm F814W}$ vs.\,$M_{\rm F275W}-M_{\rm F280N}$ and $M_{\rm F814W}$ vs.\,$M_{\rm F343N}-M_{\rm F373N}$ planes. The specific chemical composition of each isochrone is quoted in Table~\ref{tab:chimica} and corresponds to constant [(C$+$N$+$O)/Fe]. In particular, we assumed that the green and the red isochrones have [Mg/Fe]=$0.4$, while both the cyan and the blue ones are depleted in magnesium by 0.5 dex. The yellow isochrones have intermediate magnesium abundance ([Mg/Fe]=$+$0.25). 
 
 Clearly, the $M_{\rm F275W}-M_{\rm F280N}$ colour of RGB stars mostly depends on the magnesium abundance, with the Mg-poor isochrones having redder RGBs than Mg-rich isochrones. Nitrogen and helium poorly affect the $M_{\rm F275W}-M_{\rm F280N}$ colour of RGB stars. Indeed, the red and green isochrones, which have the same magnesium abundance but different nitrogen content ([N/Fe]=1.21) are superimposed to each other, similarly to the blue and cyan isochrones, which share the same value of [Mg/Fe] but have different helium mass fractions of Y=0.246 and Y=0.34, respectively.

 As illustrated in the middle panel of Figure~\ref{fig:ChM280teo}, the $M_{\rm F343N}-M_{\rm F373N}$ colour of RGB stars is mainly a tracer of the nitrogen abundance, and the RGBs moves towards red colours when the value of [N/Fe] increases. The helium abundance of the stellar populations poorly affects the $M_{\rm F343N}-M_{\rm F373N}$ colour of RGB stars, as demonstrated by the fact that the blue and cyan isochrones, which have different helium abundances and the same [N/Fe], are nearly coincident in the $M_{\rm F814W}$ vs.\,$M_{\rm F275W}-M_{\rm F280N}$ CMD.
 
 Right panel of Figure~\ref{fig:ChM280teo} shows the ChM of the RGB stars with $-1<M_{\rm F814W}<2$. 
 The vectors superimposed on the ChM represent the expected correlated changes of $\Delta_{\rm F343N,F373N}$ and $\Delta_{\rm F275W,F280N}$ when the abundances of He, C, N, O and Mg are changed one at a time. 
  Specifically, we assumed helium mass fraction variation $\Delta$\,Y=0.154 and elemental variations of $\Delta$[C/Fe]=$-$0.50, $\Delta$[N/Fe]=1.21, $\Delta$[O/Fe]=$-$0.50 and $\Delta$[Mg/Fe]=$-$0.50. 
  
 In monometallic GCs, the  $\Delta_{\rm F343N,F373N}$ quantity is nearly entirely affected by nitrogen variations whereas the value of $\Delta_{\rm F275W,F280N}$ is mostly dependent on [Mg/Fe]. 
  Star-to-star differences in He, C and O provide small variations of $\Delta_{\rm F343N,F373N}$ and $\Delta_{\rm F275W,F280N}$, that can be appreciated in the inset of Figure~\ref{fig:ChM280teo}, which is a zoom of the region of the ChM around the origin of the vectors. 
   We conclude that the $\Delta_{\rm F343N,F373N}$ vs.\,$\Delta_{\rm F275W,F280N}$ ChM is a efficient tool to identify stellar populations with different [Mg/Fe] and [N/Fe] in monometallic GCs.

\begin{table}
  \caption{Chemical composition of the five isochrones shown in Figure~\ref{fig:ChM280teo}. All isochrones have ages of 13 Gyr, [Fe/H]=$-1.8$, [$\alpha$/Fe]=0.4 and the same overall C$+$N$+$O content. }

\begin{tabular}{ c c c c c c}
\hline \hline
 Isochrone & Y & [C/Fe] & [N/Fe] & [O/Fe] & [Mg/Fe]  \\
\hline
 I1  & 0.246  &   0.00  &  0.00 &    0.40   &    0.40 \\
 I2  & 0.246  &$-$0.05  &  0.53 &    0.35   &    0.40 \\
 I3  & 0.246  &$-$0.10  &  0.93 &    0.20   &    0.25 \\
 I4  & 0.246  &$-$0.50  &  1.21 & $-$0.10   & $-$0.10 \\
 I5  & 0.340  &$-$0.50  &  1.21 & $-$0.10   & $-$0.10 \\
     \hline\hline
\end{tabular}
  \label{tab:chimica}
 \end{table}

\section{Metal-poor stellar populations}\label{sec:mpoor}

Recent work shows that GCs can be classified into two main classes of clusters with distinct photometric and spectroscopic properties.  Type I GCs exhibit single sequences of 1G and 2G stars in their ChMs and have nearly constant metal content.  Type\,II GCs are composed of multiple sequences of 1G and 2G stars in the ChM and split SGBs in CMDs made with photometry in optical bands \citep[][]{milone2015a, milone2017a}.

 \citet{marino2019a} provided the chemical tagging of multiple populations over the ChM and concluded that Type II GCs correspond to the class of `anomalous' GCs, which exhibit star-to-star variations in some heavy elements, like Fe and s-process elements \citep[e.g.][]{yong2008b, marino2009a, marino2015a, marino2018a, carretta2010a, johnson2015a}. We exploit these findings to identify a sample of metal-poor stars in $\omega$\,Cen, which is a Type II GC with extreme metallicity variations. Indeed, the fact that stars with different metallicities of Type\,II GCs occupy different regions of the ChM makes it possible to identify stellar populations with different iron abundance based on the ChM alone. 

\begin{centering} 
\begin{figure*} 
  \includegraphics[height=13.cm,trim={1.5cm 5cm 1.2cm 2.5cm},clip]{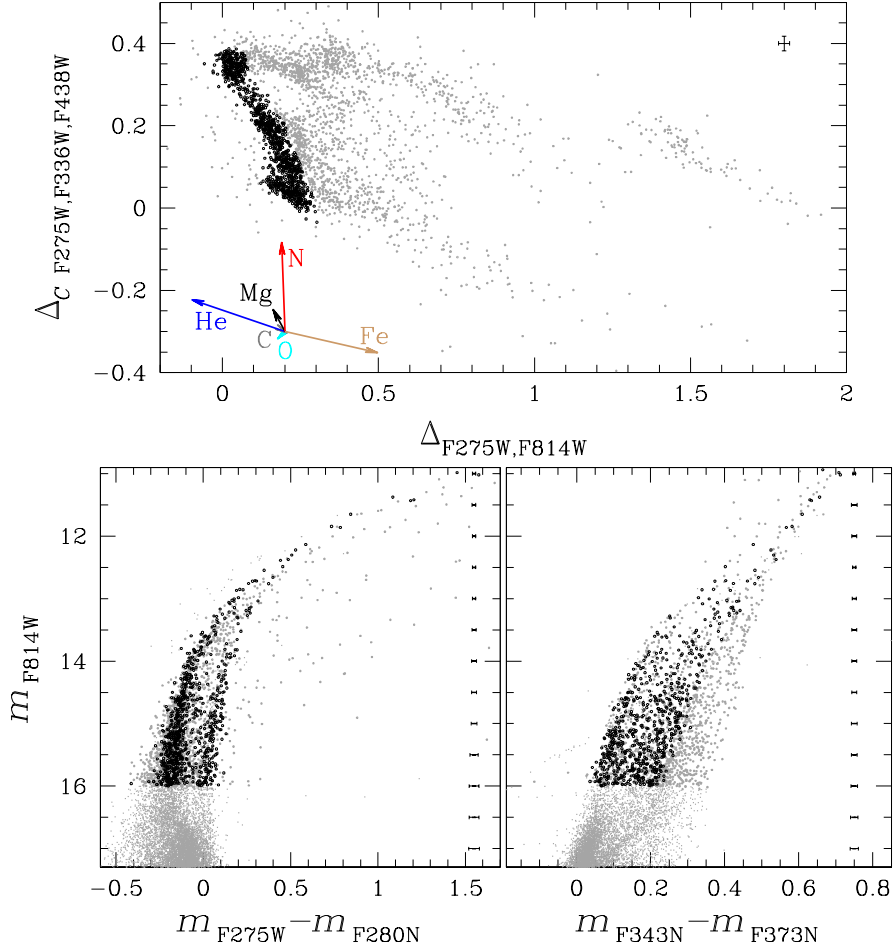}
  \caption{This figure illustrates the procedure to identify metal-poor stars in $\omega$\,Cen. the upper panel reproduces the $\Delta_{\rm {\it C} F275W,F336W,F438W}$ vs.\,$\Delta_{\rm F275W,F814W}$ ChM of $\omega$\,Cen from \citet{milone2017a}. Lower panels show the $m_{\rm F814W}$ vs.\,$m_{\rm F275W}-m_{\rm F280N}$ (left) and the $m_{\rm F814W}$ vs.\,$m_{\rm F343N}-m_{\rm F373N}$ CMD (right) of $\omega$\,Cen. Metal-poor stars are marked with black dots.
  The arrows plotted in the upper panel show the effect of changing He, C, N, O, Mg and Fe, one at a time, on $\Delta_{\rm {\it C} F275W,F336W,F438W}$ and $\Delta_{\rm F275W,F814W}$.
  We adopted an iron variation of $\Delta$[Fe/H]=$0.30$ and the same C, N, O and Mg variations of Figure~\ref{fig:ChM280teo}.   }
 \label{fig:CMDtI} 
\end{figure*} 
\end{centering}

The upper panel of Figure~\ref{fig:CMDtI} 
shows the ChM of $\omega$\,Cen where the metal-poor stars ([Fe/H]$\sim -1.8$) identified by \citet{milone2017a} are represented with black dots and coloured gray the remaining stars. 
 The sample of metal poor stars has been selected on the basis of the fact that it define the reddest RGB sequence in $m_{\rm F336W}$ vs.\,$m_{\rm F336W}-m_{\rm F814W}$ CMD, which is sensitive to stellar populations with different metallicities in Type II GCs \citep[e.g.][]{marino2011b, marino2015a, marino2018a}. The selected metal-poor stars also populate the bluest RGB sequence in CMD made with optical filters (e.g.\,$m_{\rm F438W}$ vs.\,$m_{\rm F438W}-m_{\rm F814W}$) and define a distinct sequence in the ChM.  Furthermore, the low metallicity of these stars is confirmed by direct spectroscopic measurements of the iron abundance from \citet{johnson2010a, marino2011a, mucciarelli2019a}.
Lower panels 
highlight metal-poor stars in the $m_{\rm F814W}$ vs.\,$m_{\rm F275W}-m_{\rm F280N}$ and the $m_{\rm F814W}$ vs.\,$m_{\rm F343N}-m_{\rm F373N}$ CMDs.
The RGB is clearly split in the left-panel CMD and the RGB sequence with redder $m_{\rm F275W}-m_{\rm F280N}$ colours includes $\sim$30\% of the total number of RGB stars. The colour separation between the two main RGBs is about 0.2 mag at $m_{\rm F814W}=16.0$ and decreases towards bright luminosities. The two RGBs merge together at $m_{\rm F814W}\sim12.5$. 

The discreteness of the various RGB sequences is less evident in the $m_{\rm F814W}$ vs.\,$m_{\rm F343N}-m_{\rm F373N}$ CMD (bottom-right panel of Figure~\ref{fig:CMDtI}). Metal-poor stars span a colour range of about 0.2 mag in the luminosity interval that ranges from the RGB base to $m_{\rm F814W}\sim13.5$, which is narrower than the $m_{\rm F343N}-m_{\rm F373N}$ spanned by metal rich stars. The colour spread decreases for  $m_{\rm F814W}\lesssim13.5$ and is comparable with observational errors towards the RGB tip. 

\subsection{The $\Delta_{\rm F343N,F373N}$ vs.\,$\Delta_{\rm F275W,F280N}$ chromosome map}
\begin{centering} 
\begin{figure} 
  \includegraphics[height=7.5cm,trim={.5cm 5cm .5cm 2.4cm},clip]{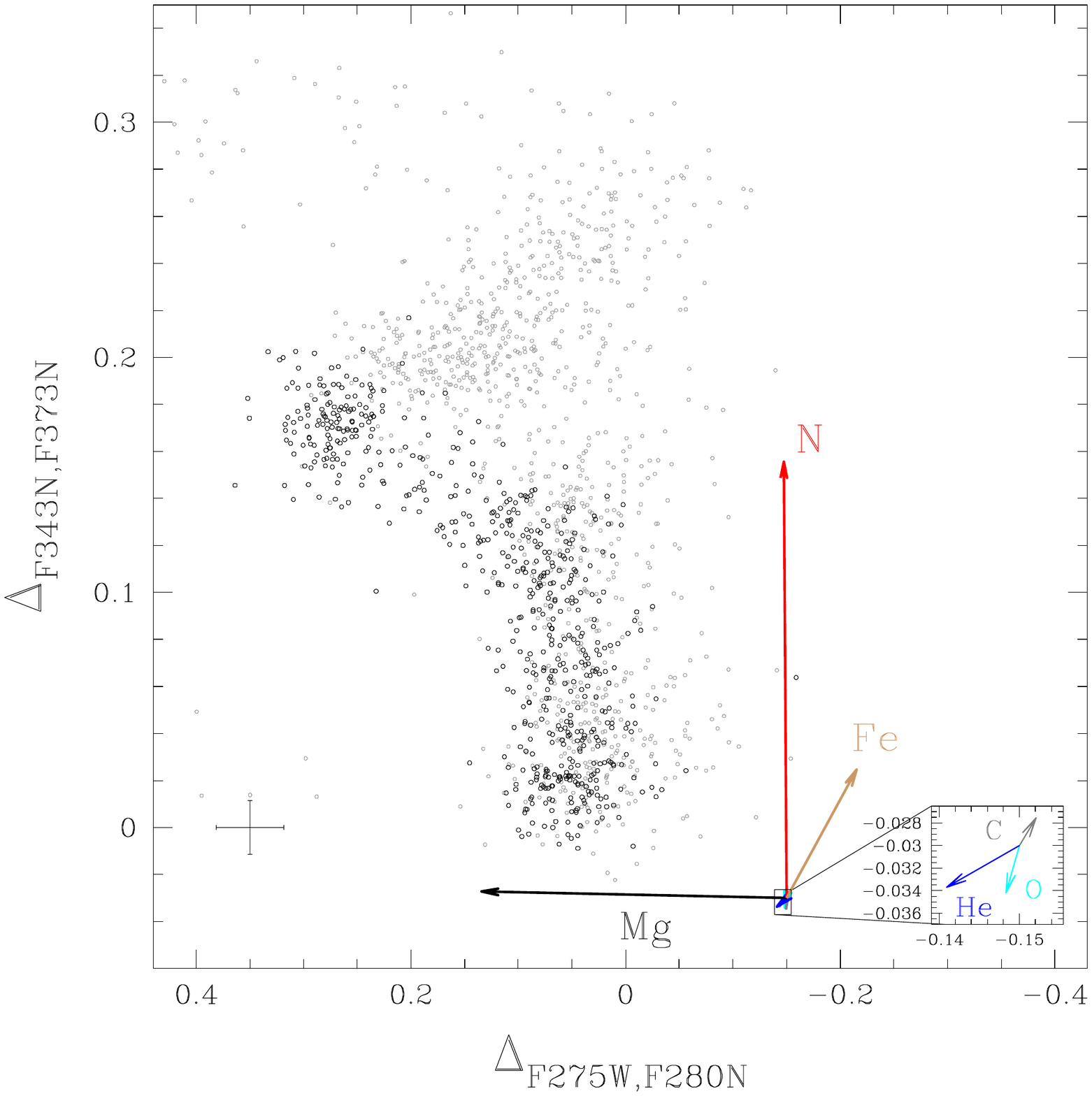}
  \caption{ $\Delta_{\rm F343N,F373N}$ vs.\,$\Delta_{\rm F275W,F280N}$ ChM of RGB stars with $13.8<m_{\rm F814W}<16.0$. Metal-poor stars are represented with black circles, while the remaining stars are coloured gray. The arrows indicate the effect of changing He, C, N, O, Mg and Fe, one at a time, on $\Delta_{\rm F343N,F373N}$ and $\Delta_{\rm F275W,F280N}$.}
 \label{fig:ChM} 
\end{figure} 
\end{centering}

The $m_{\rm F814W}$ vs.\,$m_{\rm F275W}-m_{\rm F280N}$ and $m_{\rm F814W}$ vs.\,$m_{\rm F343N}-m_{\rm F373N}$ CMDs plotted in the bottom panels of Figure~\ref{fig:CMDtI}, are used to derive the $\Delta_{\rm F343N,F373N}$ vs.\,$\Delta_{\rm F275W,F280N}$ ChM of RGB stars that we plot in Figure~\ref{fig:ChM} \citep[see][for details]{milone2015a, milone2018a, zennaro2019a}. We only included stars with $13.8<m_{\rm F814W}<16.0$, which is the magnitude interval where multiple populations are clearly visible. 

 Figure~\ref{fig:ChM} reveals that about 70\% of metal-poor stars define a vertical sequence with $\Delta_{\rm F275W,F280N} \sim 0.05$, while most of the remaining metal-poor stars are clustered around $\Delta_{\rm F275W,F280N} \sim 0.3$ and $\Delta_{\rm F343N,F373N} \sim 0.18$. 
 As demonstrated in Section~\ref{sub:NTeo}, the abscissa of this ChM is mostly sensitive to magnesium abundance.
 Hence, we conclude that about 30\% of the selected metal-poor stars in $\omega$\,Cen are significantly depleted in [Mg/Fe]. 

\begin{centering} 
\begin{figure*} 
  \includegraphics[height=10.cm,trim={0.5cm 5cm 0.5cm 7.5cm},clip]{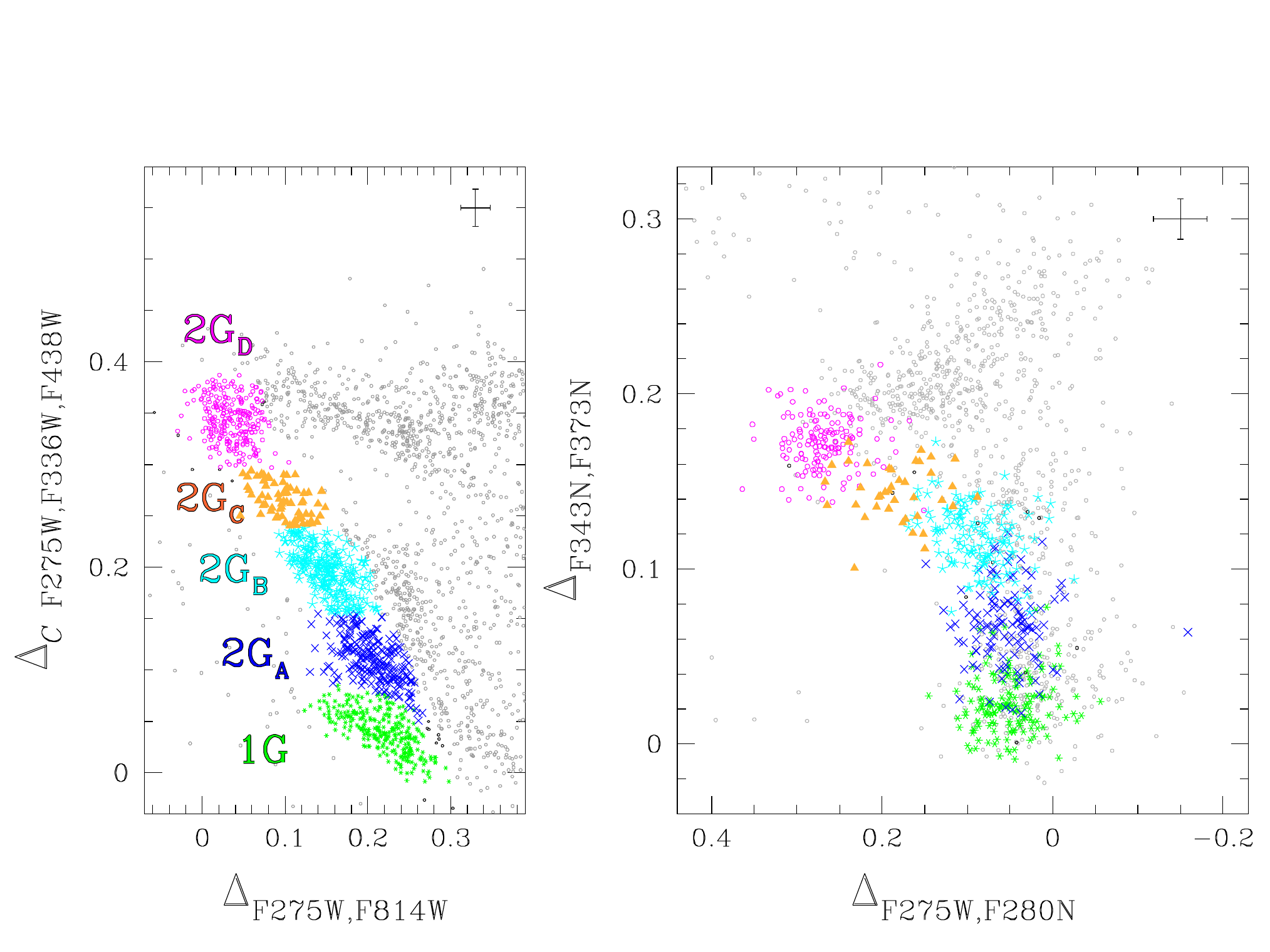}
  \caption{Zoom in of the $\Delta_{\rm {\it C} F275W,F336W,F438W}$ vs.\,$\Delta_{\rm F275W,F814W}$ ChM around the region populated by metal-poor stars (left panel) and  $\Delta_{\rm F275W,F280N}$ vs.\,$\Delta_{\rm F343N,F373N}$ ChM (lower panel) of RGB stars of $\omega$\,Cen.
  1G stars are represented with green symbols, while the groups of 2G$_{\rm A--D}$ metal-poor stars are coloured blue, cyan, orange and magenta, respectively. Gray dots represent the remaining RGB stars.
   }
 \label{fig:popst1} 
\end{figure*} 
\end{centering}

Figure~\ref{fig:popst1} compares the classical ChM of $\omega$ Cen from \citet{milone2017a} with the $\Delta_{\rm F275W,F280N}$ vs.\,$\Delta_{\rm F343N,F373N}$ ChM introduced in this paper. We used green colours to mark the 1G stars defined by Milone and collaborators, which contains 25.9$\pm$1.3\% of stars. We also identified four main groups of 2G stars with different values of  $\Delta_{\rm {\it C} F275W,F336W,F438W}$, that we called 2G$_{\rm A--D}$ and include 20.7$\pm$1.2, 23.2$\pm$1.3, 6.3$\pm$0.7 and 23.8$\pm$1.3\% of metal-poor stars, respectively. Their stars are represented in Figure~\ref{fig:popst1} with blue, cyan, orange and magenta colours, respectively. 
We find that 2G stars have, on average, higher values of $\Delta_{\rm F275W,F280N}$ and $\Delta_{\rm F343N,F373N}$  than 1G stars. 
 2G$_{\rm A}$ and 2G$_{\rm B}$ stars share almost the same range of $\Delta_{\rm F275W,F280N}$ as 1G stars, but have higher $\Delta_{\rm F343N,F373N}$ values than the 1G. 
  2G$_{\rm D}$ stars exhibit significantly higher values of $\Delta_{\rm F275W,F280N}$ and  $\Delta_{\rm F343N,F373N}$ than the remaining RGB stars, while the poorly-populated group of 2G$_{\rm C}$ stars exhibits intermediate values of $\Delta_{\rm F275W,F280N}$ and $\Delta_{\rm F343N,F373N}$. 
  
  Since  $\Delta_{\rm F275W,F280N}$ quantity is mostly affected by magnesium variation, we conclude that 1G stars share similar [Mg/Fe] as 2G$_{\rm A}$ and 2G$_{\rm B}$ stars and correspond to the populations with [Mg/Fe]$\sim$0.5 identified by \citet{norris1995a}, while the remaining stars are significantly depleted in magnesium with respect to the bulk of $\omega$\,Cen stars, with  2G$_{\rm D}$ stars having the lowest magnesium content. 
  The fact that in monometallic GCs the $\Delta_{\rm F343N,F373N}$ quantity is mostly sensitive to nitrogen variations, indicates that the nitrogen abundance increases when we move from the 1G, which has the lowest value of [N/Fe], to 2G$_{\rm D}$. In the next section, we will exploit photometry in 36 filters to 
  constrain the chemical composition of the metal-poor stellar populations in $\omega$\,Cen.
   
\subsection{The chemical composition of stellar populations}\label{sub:chimica}
To infer the relative abundances of He, C, N, O, and Mg between 2G$_{\rm A--D}$ and 1G stars, we extended  the method by \citet{milone2012b, milone2018a} and \citet{lagioia2018a, lagioia2019a} to the sample of metal-poor stars of $\omega$\,Cen. 
Briefly, we derived the fiducial lines of  2G$_{\rm A--D}$ and 1G stars in the $m_{\rm F814W}$ vs.\,$m_{\rm X}-m_{\rm F814W}$ CMD in the luminosity interval between $m_{\rm F814W}=13.8$ and $m_{\rm F814W}=16.0$, where X corresponds to each of the 36 filters used in this paper and listed in Section~\ref{sec:data}.  
 To do this, we divided the RGB into F814W magnitude intervals of
size $\delta m$=0.25 mag which are defined over a grid of points spaced by magnitude $\delta m/2$ bins of size. For each bin we calculated the median
  $m_{\rm F814W}$ magnitude and $m_{\rm X}-m_{\rm F814W}$ color and linearly interpolated these median points.

\begin{centering} 
\begin{figure*} 
  \includegraphics[height=6.cm,trim={0.5cm 5cm 2.25cm 8.9cm},clip]{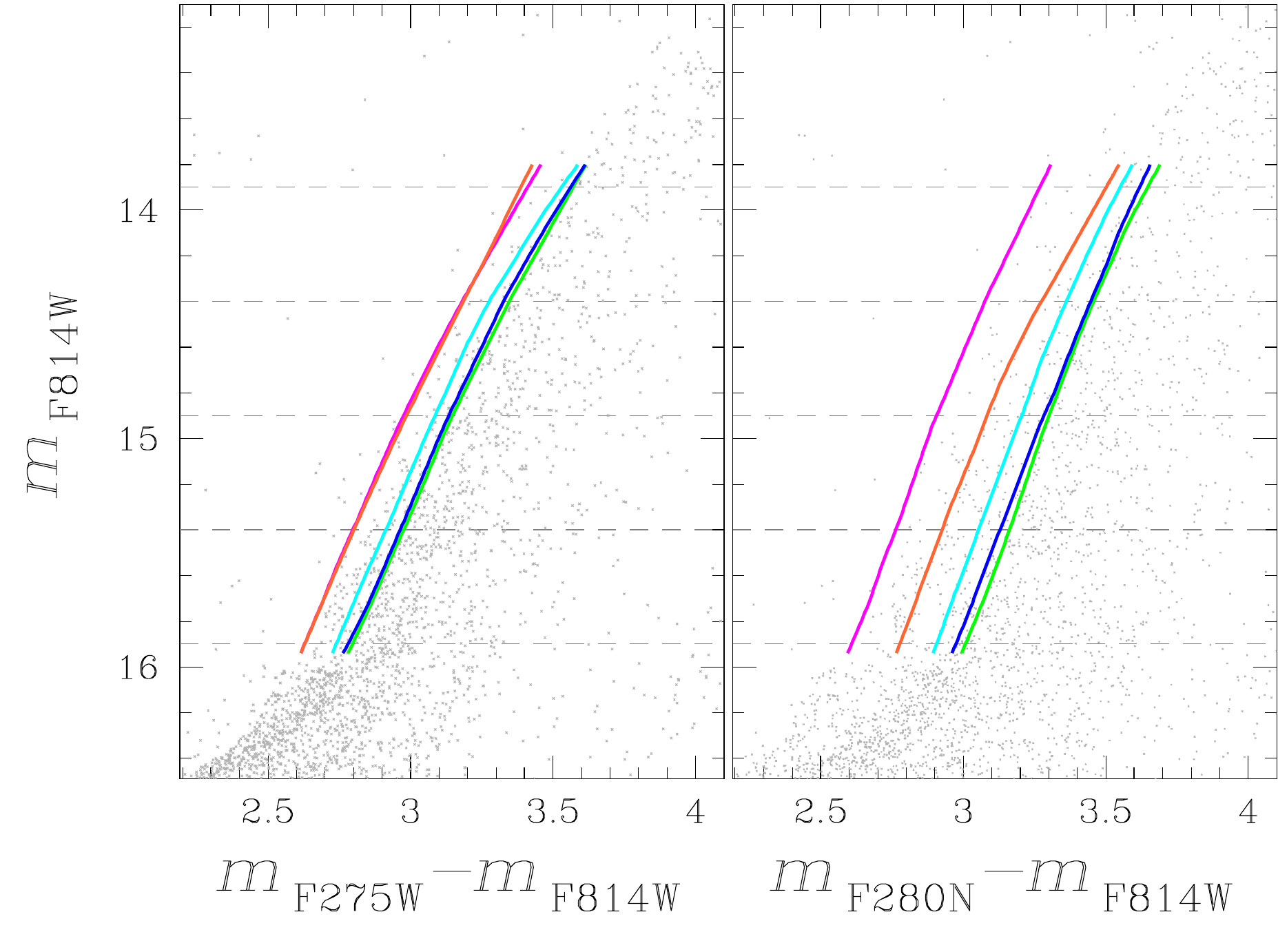}
  \includegraphics[height=6.cm,trim={3.15cm 5cm 0.5cm 8.9cm},clip]{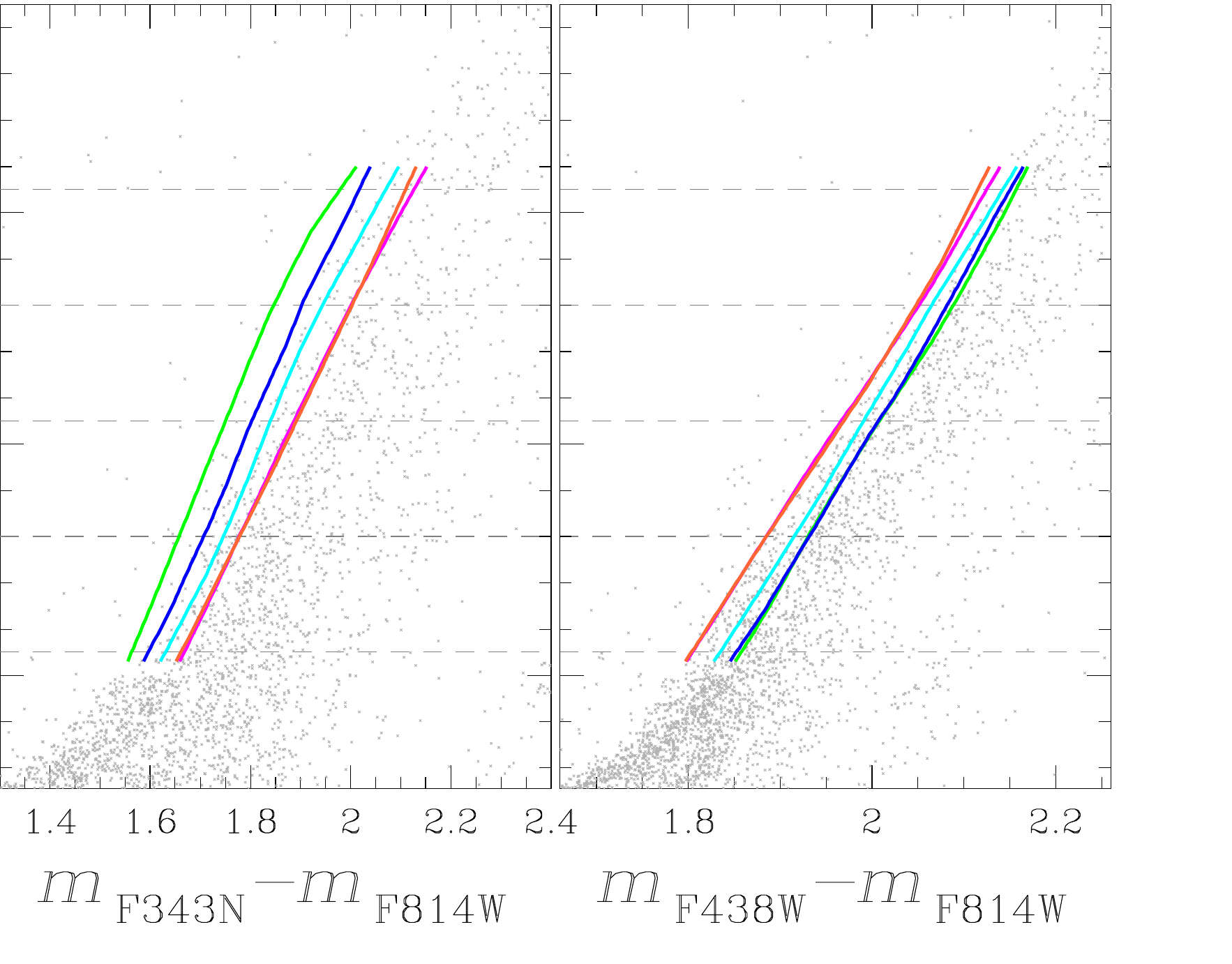}
  \includegraphics[height=7.5cm,trim={.5cm 5cm 0.5cm 2.9cm},clip]{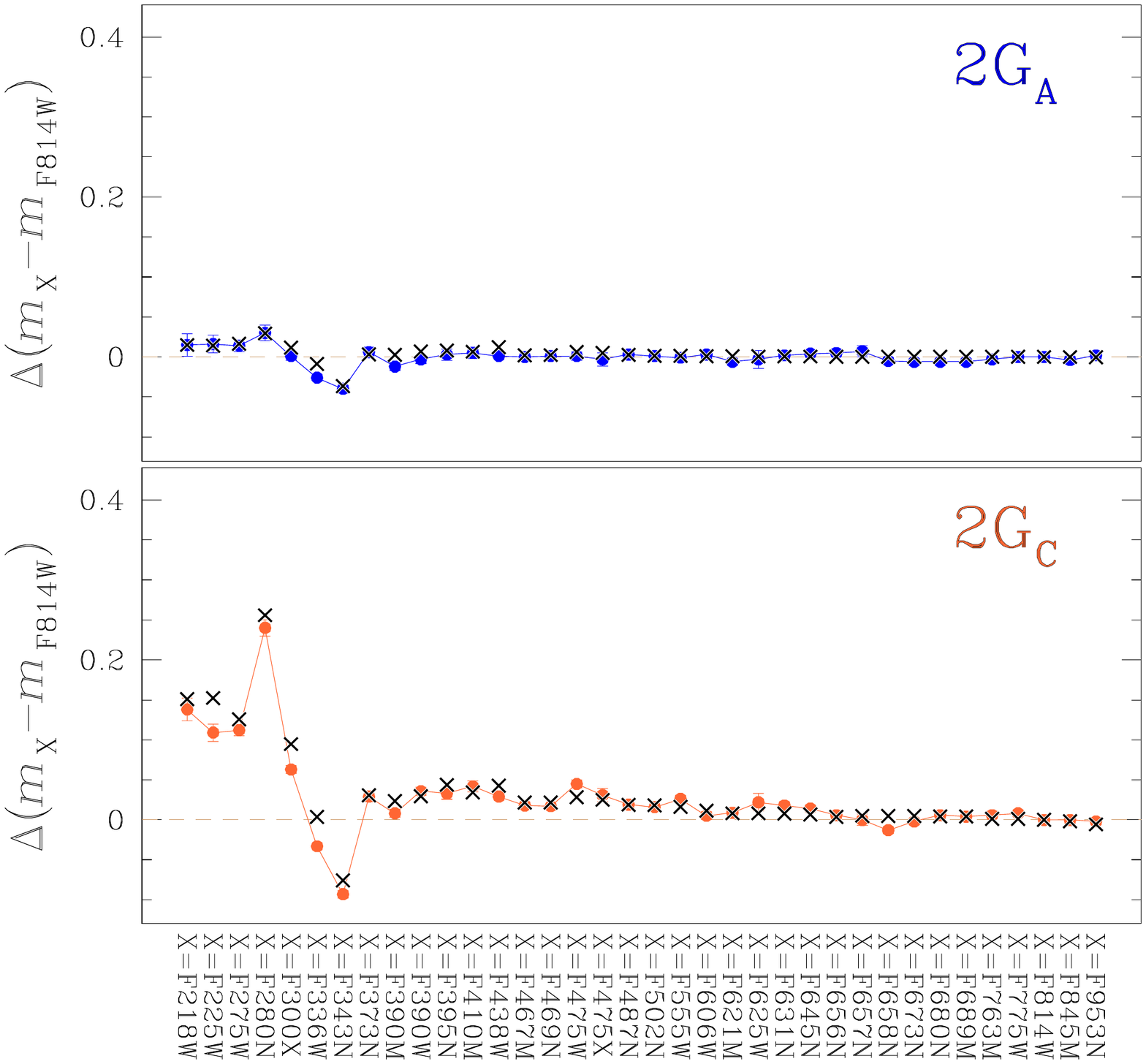}
  \includegraphics[height=7.5cm,trim={3.15cm 5cm 0.5cm 2.9cm},clip]{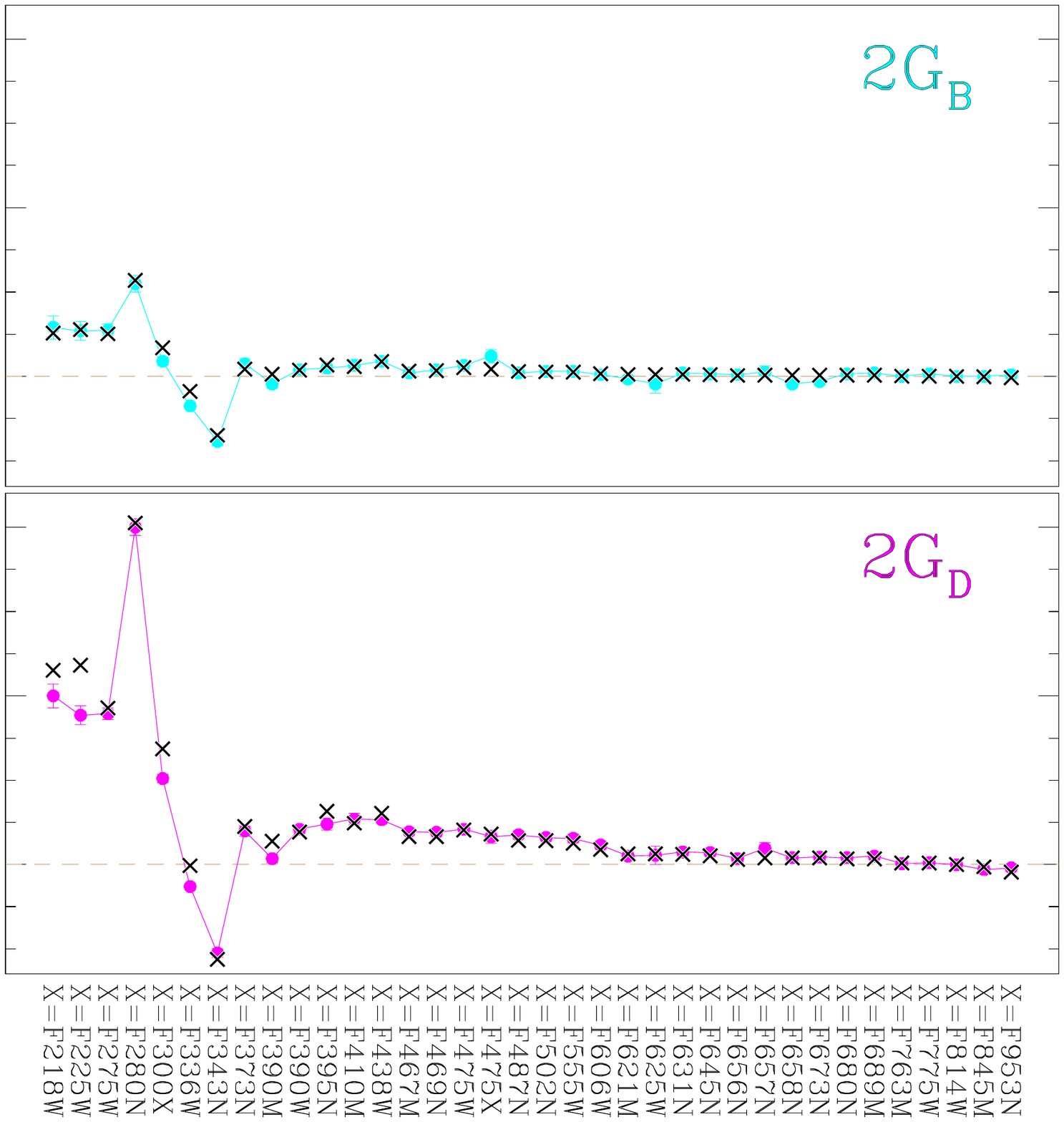}
 
  \caption{Upper panels show the $m_{\rm F814W}$ vs.\,$m_{\rm X}-m_{\rm F814W}$ CMDs for RGB stars, where X=F275W, F280N, F343N and F438W. The fiducial lines of the 1G and 2G$_{\rm D}$ populations identified in the paper are coloured green and magenta, respectively, while the dashed lines mark the five values of $m_{\rm F814W}^{\rm CUT}$ used to estimate the relative chemical compositions of the various stellar populations. Lower panels show the $m_{\rm X}-m_{\rm F814W}$ colour differences between the fiducials of 2G$_{\rm A--D}$ stars and the fiducial of 1G stars for the 36 X filters calculated at $m_{\rm F814W}^{\rm CUT}=15.9$.  Observations are represented with coloured points while black crosses correspond to the best-fit models.  The procedure, illustrated in the lower panels for $m_{\rm F814W}^{\rm CUT}=15.9$ has been extended to the other four values of $m_{\rm F814W}^{\rm CUT}$.}
 \label{fig:fidu} 
\end{figure*} 
\end{centering}

As an example, in the upper panels of Figure~\ref{fig:fidu} we plot $m_{\rm F814W}$ against $m_{\rm X}-m_{\rm F814W}$ for RGB stars of $\omega$\,Cen, where X=F275W, F280N, F343N and F438W and use green, blue, cyan, orange and magenta colours to represent the fiducials of the 1G, 2G$_{\rm A--D}$.  

We defined five reference magnitudes along the RGB, namely $m^{\rm ref}_{\rm F814W}=15.9, 15.4, 14.9, 14.4$ and 13.9  to derive five estimates of the chemical composition of stellar populations. These five magnitude levels are marked with dashed horizontal lines in the upper panels of Figure~\ref{fig:fidu}. 
For each value of $m^{\rm ref}_{\rm F814W}$ we measured the colour difference between the fiducial of  2G$_{\rm A--D}$ and the fiducial of 1G stars ($\Delta$($m_{\rm X}-m_{\rm F814W}$)). 
 
\begin{centering} 
\begin{figure} 
  \includegraphics[height=9.cm,trim={0.0cm 6cm 0cm 0cm},clip]{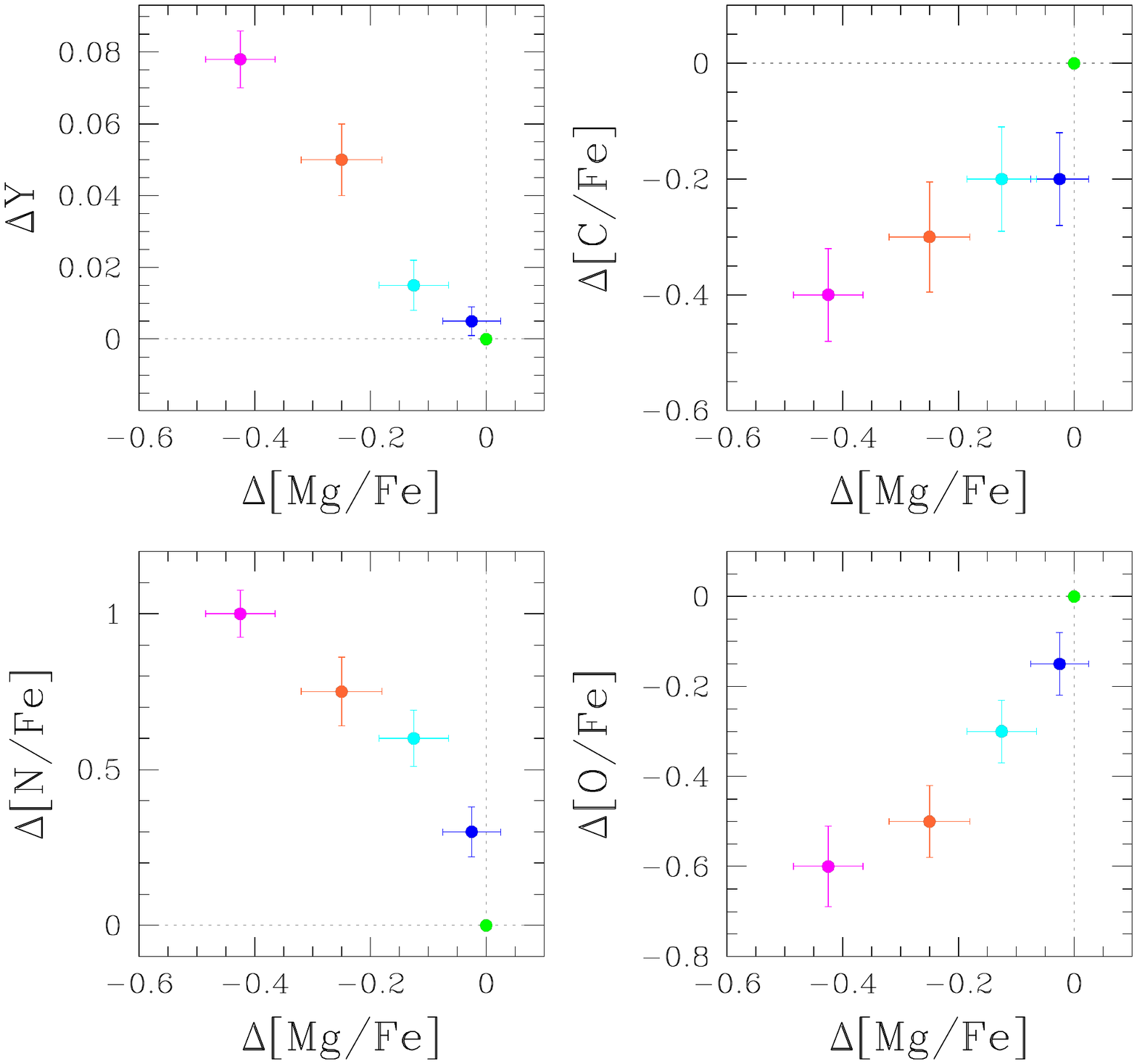}
  \caption{The abundances of helium, carbon, nitrogen and oxygen of the five studied stellar populations relative to the 1G abundances are plotted against the  relative magnesium content. Green, blue, cyan, orange and magenta colours represent 1G and 2G$_{A--D}$ stellar populations, respectively.}
 \label{fig:chimica} 
\end{figure} 
\end{centering} 
We exploited the isochrones by \citet{dotter2008a} that provide the best match with the observed $m_{\rm F814W}$ vs.\,$m_{\rm F606W}-m_{\rm F814W}$ CMD to estimate the gravities and the effective temperatures of 1G stars corresponding to the five reference points. To do this, we assumed [Fe/H]=$-$1.8, which is similar to the iron abundance inferred by \citet{marino2019a} for metal-poor stars of $\omega$\,Cen, [$\alpha$/Fe]=0.4. We also adopted age of 13.0 Gyr, distance modulus ($m-M$)$_{\rm 0}=13.58$ and reddening E(B$-$V)=0.15, which are similar to the quantities provided by \citet[][updated as in 2010]{harris1996a} and \citet{dotter2010a}. 

To constrain the chemical composition of 2G$_{\rm A--D}$ stars  from the colors calculated at  $m^{\rm ref}_{\rm F814W}=15.9$, we computed a grid of synthetic spectra with different abundances of He, C, N, O and Mg, that we compare with a synthetic spectrum corresponding to the 1G used as reference. The latter has the values of gravity and effective temperature inferred from the best-fit isochrone  for $m^{\rm ref}_{\rm F814W}=15.9$, Y=0.246, [C/Fe]=0.0, [N/Fe]=0.0, [O/Fe]=0.4 and [Mg/Fe]=0.4. 
 We assumed for synthetic spectra a set of values for [C/Fe] that range from $-$0.5 to 0.1, [N/Fe] between 0.0 and 1.5, [O/Fe] that ranges from $-0.6$ to $0.5$ and [Mg/Fe] from $-0.2$ to $0.5$. We used steps of 0.05 dex for all elements.
The helium mass fractions of the comparison spectra range from Y=0.246 to
0.400 in steps of 0.001 and the effective parameters  of helium-rich isochrones are taken from the corresponding isochrones from \citet{dotter2008a}. The spectra are computed by using the computer programs ATLAS 12 and Synthe \citep{sbordone2004a, sbordone2007a, kurucz2005a, castelli2005a}.

We convoluted each spectrum with the transmission curves of the 36 WFC3/UVIS and ACS/WFC filters available for $\omega$\,Centauri and derived the colour difference between 2G$_{\rm A--D}$ and 1G stars corresponding to each reference point. The best estimates of Y, C, N, O, and Mg of 2G$_{\rm A--D}$ stars,  for $m^{\rm ref}_{\rm F814W}=15.9$ are given by the elemental abundances of the comparison synthetic spectrum that provides the best fit with observed colour differences. 
The best fit is derived by means of $\chi$-square minimization, that is calculated by accounting for the uncertainties on color determinations and on the sensitivity of the colors to the abundance of a given element, as inferred from synthetic spectra \citep[see][]{dodge2008a}. In the following cases, we excluded some filters to derive the abundances of certain elements. Specifically, we constrained the helium content from filters that are redder than F438W  as they are poorly affected by the effect of C, N, O and Mg on stellar atmosphere. We used the F336W and F343N filters to infer the content of nitrogen alone, as they are largely affected by the abundance of this element. Similarly, we used F280N to derive the magnesium abundance alone.
An example is provided in the lower panels of Figure~\ref{fig:fidu}, where we compare the colour differences between the fiducials of 2G$_{\rm A--D}$ stars and the fiducial of the 1G calculated at $m_{\rm F814W}^{\rm cut}=15.9$ (coloured points) with the colour differences from the best-fit comparison spectra (black crosses).  

 This procedure, discussed and illustrated for $m_{\rm F814W}^{\rm cut}=15.9$, is extended to the other four reference magnitudes and provides five determinations of He, C, N, O and Mg of population 2G$_{\rm A--D}$-stars relative to the 1G.
 The average abundances are listed in Table~\ref{tab:abb}. 
 
\begin{table*}
  \caption{Average abundances of He, C, N, O and Mg of population 2G$_{\rm A--D}$-stars relative to the 1G.}
\begin{tabular}{ c c c c c c}
\hline \hline
 Population & $\Delta$Y & $\Delta$[C/Fe] & $\Delta$[N/Fe] & $\Delta$[O/Fe] & $\Delta$[Mg/Fe]  \\
\hline
 2G$_{\rm A}$  & 0.005$\pm$0.004  &  $-$0.20$\pm$0.08  & 0.31$\pm$0.08 &   $-$0.15$\pm$0.08  & $-$0.03$\pm$0.04 \\
 2G$_{\rm B}$  & 0.016$\pm$0.007  &  $-$0.20$\pm$0.09  & 0.62$\pm$0.09 &   $-$0.30$\pm$0.07  & $-$0.13$\pm$0.06 \\
 2G$_{\rm C}$  & 0.051$\pm$0.010  &  $-$0.32$\pm$0.11  & 0.75$\pm$0.10 &   $-$0.50$\pm$0.09  & $-$0.25$\pm$0.07 \\
 2G$_{\rm D}$  & 0.081$\pm$0.007  &  $-$0.42$\pm$0.08  & 1.02$\pm$0.07 &   $-$0.60$\pm$0.09  & $-$0.44$\pm$0.06 \\
     \hline\hline
\end{tabular}
  \label{tab:abb}
 \end{table*}

 To test the robustness of the results  and estimate the uncertainties, we repeated the procedure described above  on 1,000 simulated $m_{\rm F814W}$ vs.\,$m_{\rm X}-m_{\rm F814W}$ CMDs with fixed light-element variations and the same number of stars and observational errors as our observations of $\omega$\,Centauri. We find that our procedure correctly recovers the input abundances  and provides the uncertainties listed in Table~\ref{tab:abb}. 

We find that 2G$_{\rm A}$ and 1G stars have similar magnesium content, while 2G$_{\rm D}$ stars exhibit extreme magnesium depletion by $\Delta$[Mg/Fe]$\sim$0.45 dex, with respect to the 1G. 2G$_{\rm B}$ and 2G$_{\rm C}$ exhibit intermediate magnesium abundances.
As shown in Figure~\ref{fig:chimica}, the abundances of helium and nitrogen anti-correlate with magnesium content, while carbon and oxygen correlate with [Mg/Fe]. 
2G$_{\rm D}$ stars, which exhibit the largest abundance variations, are enhanced in [N/Fe] by $\sim$1.0 dex and in helium mass fraction by $\sim$0.08, with respect to the 1G. Moreover, 2G$_{\rm D}$ stars are depleted in carbon and oxygen by $\sim$0.4 and $\sim$0.6 dex, when compared with the 1G. 
 2G$_{\rm A--C}$ stars exhibit intermediate abundances of He, C, N and O.

\begin{centering}  
\begin{figure*} 
  \includegraphics[height=13.cm,trim={1.5cm 5cm 1.2cm 2.5cm},clip]{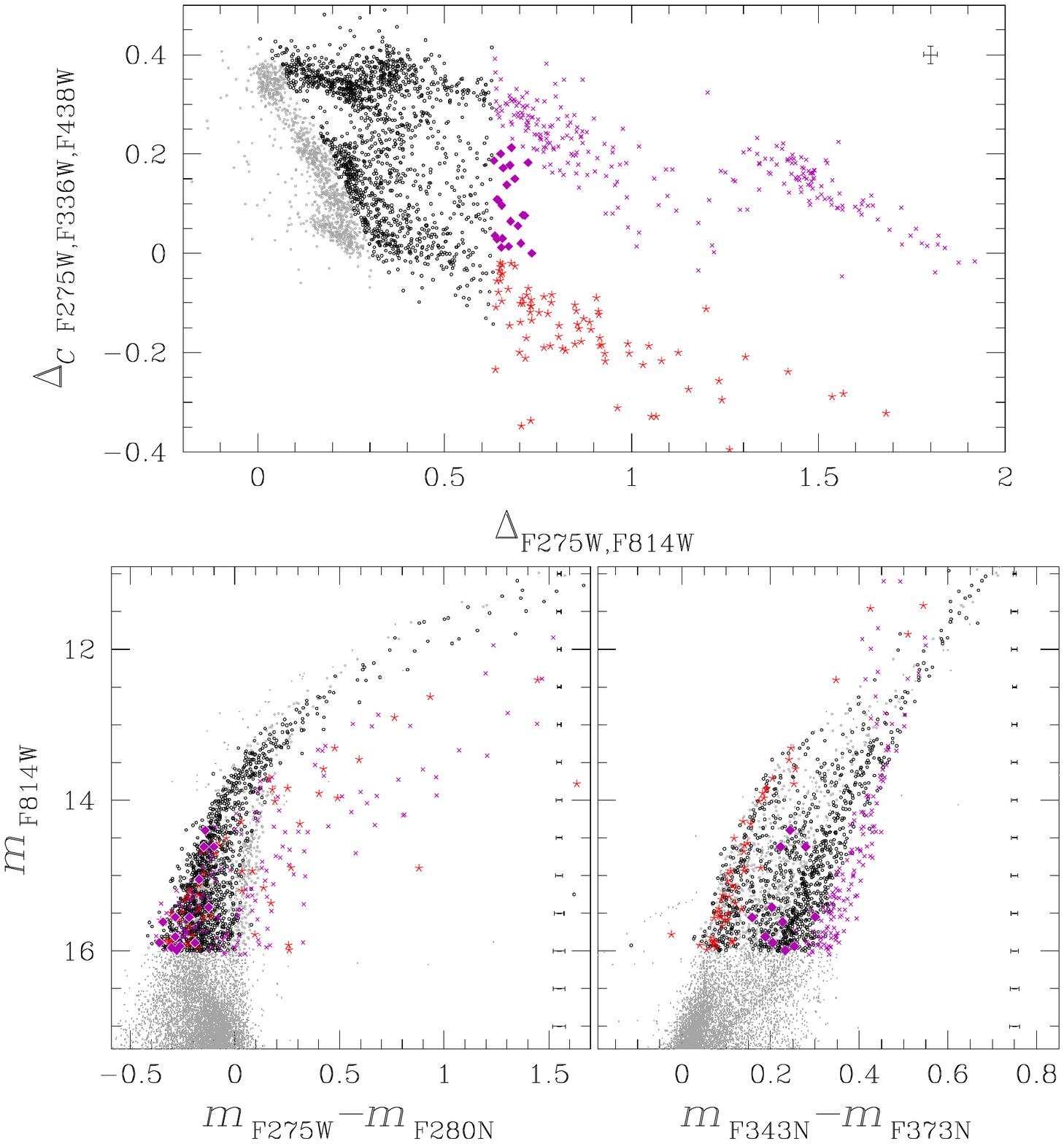}
  \caption{  $\Delta_{\rm {\it C} F275W,F336W,F438W}$ vs.\,$\Delta_{\rm F275W,F814W}$ ChM (upper panel), $m_{\rm F814W}$ vs.\,$m_{\rm F275W}-m_{\rm F280N}$ (left) (lower-left panel) and $m_{\rm F814W}$ vs.\,$m_{\rm F343N}-m_{\rm F373N}$ CMD (lower-right panel) of $\omega$\,Cen. Metal-rich stars with $\Delta_{\rm F275W,F814W}<0.65$ are coloured black. The remaining metal-rich stars of the lower- middle- and upper stream are represented with red starred symbols, purple diamonds and purple crosses, respectively.  }
 \label{fig:ano1} 
\end{figure*} 
\end{centering} 

\section{Metal-rich stellar populations}\label{sec:mrich}
In this section, we investigate the sample of metal-rich stars identified by \citet{milone2017a} that we highlight in the ChM and CMDs of Figure~\ref{fig:ano1}. Specifically, we used red and purple symbols to represent the metal rich stars with $\Delta_{\rm F275W,F814W}>0.6$, which comprise stars with [Fe/H]$\gtrsim -1.0$ \citep[][]{marino2019a}, and we coloured black the remaining metal-rich stars (hereafter metal-intermediate sample).  

The RGBs of metal-rich stars clearly exhibit less-steep slopes than those of metal-poor and metal-intermediate stars in the $m_{\rm F814W}$ vs.\,$m_{\rm F275W}-m_{\rm F280N}$, in qualitative agreement with the isochrones plotted in Figure~\ref{fig:iso}. 
 Metal intermediate stars with $m_{\rm F814W}>13.8$ span a smaller range of $m_{\rm F275W}-m_{\rm F280N}$ than metal poor stars with similar F814W magnitude. Moreover, there is no clear evidence for a bimodal $m_{\rm F275W}-m_{\rm F280N}$ distribution as observed for metal-poor stars.
 
 Based on chemical abundances of stars in the ChM, \citet{marino2019a} identified the lower-, middle-, and upper-streams that are composed of N-poor, N-intermediate and N-rich stars, respectively. 
 The sample of metal-intermediate stars with $m_{\rm F814W} \gtrsim 13.5$ shown in Figure~\ref{fig:ano1} span an interval of $\sim$0.1 mag in $m_{\rm F343N}-m_{\rm F373N}$ and is composed of three main RGBs. The RGB with the reddest $m_{\rm F343N}-m_{\rm F373N}$ colour is the most-populated one and is composed of stars of the upper stream of $\omega$ Cen. The bluest and the middle RGBs comprise  stars that populated the lower and the middle stream. 
 
 Stars in the lower, middle and upper stream with $\Delta_{\rm F275W,F814W}>0.65$ are represented with red starred symbols, purple diamonds and purple crosses, respectively. 
 Most of these metal-rich stars are distributed into two distinct sequences that define the bluest and the reddest boundaries of the low RGB in the $m_{\rm F814W}$ vs.\,$m_{\rm F343N}-m_{\rm F373N}$ CMD, and are composed of stars of the lower and upper stream, respectively. Middle-stream stars exhibit intermediate colours.
 All metal-rich stars exhibit bluer $m_{\rm F343N}-m_{\rm F373N}$ colours than the remaining RGB stars at magnitudes brighter than $m_{\rm F814W} \sim 13.0$, as expected from isochrones with different metallicities (see Figure~\ref{fig:iso}).

\begin{centering} 
\begin{figure*} 
  \includegraphics[height=13.cm,trim={1.0cm 5cm 1.cm 2.5cm},clip]{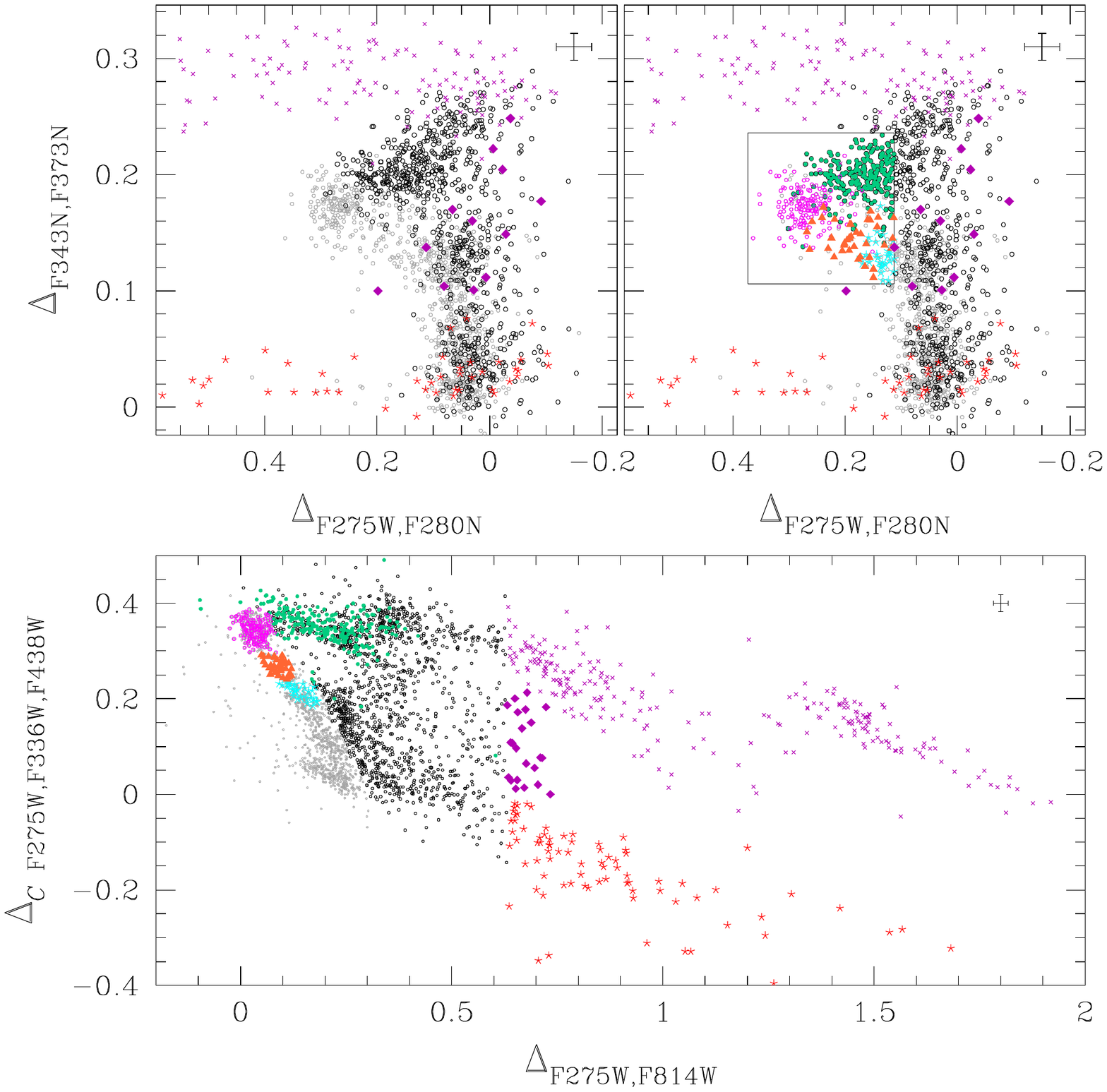}
  \caption{\textit{Upper-left panels.} Reproductions of the $\Delta_{\rm F343N,F373N}$ vs.\,$\Delta_{\rm F275W,F280N}$ ChM introduced in Figure~\ref{fig:ChM}. We used black colours to represent the metal-intermediate stars defined in Figure~\ref{fig:ano1} while metal-rich stars that belong to the upper, middle and lower stream are represented with purple crosses, purple diamonds and red starred symbols, respectively.  The remaining RGB stars are coloured gray. The box superimposed on the ChM plotted in the upper-right panel encloses stars with depleted magnesium. 
    \textit{Lower panel.} $\Delta_{\rm {\it C} F275W,F336W,F438W}$ vs.\,$\Delta_{\rm F275W,F814W}$ ChM.
     The 2G$_{\rm B}$, 2G$_{\rm C}$, and 2G$_{\rm D}$ metal-poor stars within the box shown in the upper-left panel, are coloured cyan, orange and magenta, respectively, while candidate Mg-poor metal-intermediate stars marked with aqua symbols in the upper-left and lower panels.    }
 \label{fig:ano2} 
\end{figure*} 
\end{centering} 

 The fact that metal-intermediate and metal-rich stars of $\omega$\,Cen span a wide metallicity  interval makes it challenging to estimate their relative abundances of He, C, N, O and Mg. 
Indeed, the method adopted in Section~\ref{sub:chimica} for metal-poor stars is based on  the comparison between the observed colors of multiple populations and the colors  derived by synthetic spectra with appropriate chemical composition.    Metallicity variations, including iron abundance and overall C$+$N$+$O abundances, provide significant changes on stellar structure \citep[][]{sbordone2011a}, thus affecting the atmospheric parameters of synthetic spectra.  Nevertheless, metallicity is poorly constrained for the various stellar populations of $\omega$\,Cen \citep[][]{marino2019a}. 

For this reason, we will only provide a qualitative discussion on the chemical composition of metal-intermediate and metal-rich stellar populations, which are strongly enhanced in [Fe/H] and [(C$+$N$+$O)/Fe] with respect to metal-poor stars \citep[ e.g.][]{johnson2010a, marino2011a, marino2012a}.
 
To do this, we reproduce in the upper-left panel of Figure~\ref{fig:ano2} the $\Delta_{\rm F343N,F373N}$ vs.\,$\Delta_{\rm F275W,F280N}$ ChM and highlight metal-intermediate and metal-rich stars with black and purple+red symbols, respectively. 
Most metal-rich stars define two horizontal sequences with nearly constant $\Delta_{\rm F343N,F373N} \sim 0.0$ (populated by lower-stream stars) and one with $\Delta_{\rm F343N,F373N} \sim 0.3$ that hosts upper-stream stars. Middle-stream metal-rich stars exhibit intermediate $\Delta_{\rm F343N,F373N}$ values.

Metal-intermediate stars define three main stellar blobs with $\Delta_{\rm F275W,F280N} \sim 0.05$ but different values of $\Delta_{\rm F343N,F373N} \sim 0.05, 0.13$ and 0.20 that we indicate as M-I$_{\rm A}$, M-I$_{\rm B}$ and M-I$_{\rm C}$, respectively, and are composed of stars in the lower, middle and upper stream.   
 A fourth group of metal-intermediate stars (M-I$_{\rm D}$) is distributed around ($\Delta_{\rm F275W,F280N}:\Delta_{\rm F343N,F373N}$) $\sim$ (0.15:0.20).  
 
 To highlight metal-poor and metal-intermediate stars that, based on their position in the $\Delta_{\rm F343N,F373N}$ vs.\,$\Delta_{\rm F275W,F280N}$ ChM, are strongly depleted in magnesium with respect to the majority of $\omega$\,Cen stars, we draw the gray box in the upper-right panel of Figure~\ref{fig:ano2}.
 This box includes the metal-poor populations 2G$_{\rm C}$ and 2G$_{\rm D}$ that we represented with magenta and orange points, respectively, plus a few population 2G$_{\rm B}$ stars coloured cyan.  
 The evidence that the M-I$_{\rm D}$ stars in the box (aqua dots) have larger values of $\Delta_{\rm F275W,F280N}$ than the Mg-rich metal-poor stars demonstrate that they have low magnesium content.
 
 The $\Delta_{\rm {\it C} F275W,F336W,F438W}$ vs.\,$\Delta_{\rm F275W,F814W}$ ChM plotted in the lower panel of Figure~\ref{fig:ano2} reveals that metal-intermediate Mg-poor stars belong to the upper stream. Hence, as shown by  \citet{marino2019a}, they have enhanced He, N and Na and depleted C and O.

\section{Summary and discussion}\label{sec:summary}
We exploited high-precision photometry in the F275W, F280N, F343N and F373N bands of {\it HST} of $\omega$\,Centauri, to introduce a ChM that is able to identify stellar populations with different Mg and N abundances along the RGB. 
 Based on synthetic spectra with appropriate light-element abundances, we demonstrated that the abscissa of this new ChM, $\Delta_{\rm F275W, F280N}$, is mostly sensitive to the relative content of magnesium of the stellar populations of monometallic GCs and is poorly affected by star-to-star variations of He, C, N, and O. Its ordinate, $\Delta_{\rm F343N, F373N}$, maximises the separation between stellar populations with different nitrogen and is negligibly affected by variation in the other light elements involved in the multiple-population phenomenon, including He, C, O and Mg.
In GCs with iron variations, 
both axes of the $\Delta_{\rm F343N, F373N}$ vs.\,$\Delta_{\rm F275W, F280N}$ are affected by changes in metallicity. 

To constrain the chemical composition of stellar populations in $\omega$\,Cen, we first considered the sample of metal-poor stars identified by \citet{milone2017a} on the   $\Delta_{\rm {\it C} F275W,F336W,F438W}$ vs.\,$\Delta_{\rm F275W,F814W}$ ChM, which comprises stars with nearly homogeneous [Fe/H] \citep{marino2019a}.
 We separated 1G and 2G stars as in \citet{milone2017a} and selected four main groups of 2G$_{\rm A}$--2G$_{\rm D}$ stars that span different intervals of $\Delta_{\rm {\it C} F275W,F336W,F438W}$.
 1G stars and 2G$_{\rm A}$--2G$_{\rm D}$ stars include 25.9$\pm$1.3\%, 20.7$\pm$1.2\%, 23.2$\pm$1.3\%, 6.3$\pm$0.7\% and 23.8$\pm$1.3\%, respectively, of the total number of studied metal-poor RGB stars.

As expected, the average $\Delta_{\rm F343N, F373N}$ increases from 1G- to 2G$_{\rm D}$-stars. Indeed, both $\Delta_{\rm F343N, F373N}$ and $\Delta_{\rm {\it C} F275W,F336W,F438W}$ strongly depend on the nitrogen abundance.
The fact that population 2G$_{\rm A}$ exhibits similar values of $\Delta_{\rm F275W,F280N}$ as 1G stars, shows that these populations share similar magnesium abundances. The high values of $\Delta_{\rm F275W, F280N}$ of 2G$_{\rm D}$ stars are indicative of their high depletion in [Mg/Fe], while 2G$_{\rm C}$ and and 2G$_{\rm B}$  stars have intermediate magnesium abundances. 

We exploited photometry collected through 36 filters of ACS/WFC and WFC3/UVIS on board {\it HST} and  compared the relative colours of the various populations with the colours derived from synthetic spectra with appropriate chemical compositions.    
 We thus inferred the abundances of He, C, N, O and Mg of 2G$_{\rm A--D}$ stars relative to the 1G. We find that 2G$_{\rm A}$ stars have nearly the same magnesium abundance as the 1G, while 2G$_{\rm B}$, 2G$_{\rm C}$ and 2G$_{\rm D}$ stars are depleted in magnesium by $\sim$0.15, $\sim$0.25 and $\sim$0.45 dex, respectively.  
 
 All 2G stars are more helium- and nitrogen-rich than 1G stars, with 2G$_{\rm D}$ stars having the highest enhancement of both elements ($\Delta$Y$\sim$0.08 and $\Delta$[N/Fe]$\sim$1.0). The second generations have lower content of carbon and oxygen than the 1G, with 2G$_{\rm D}$ stars having extreme  contents of these elements. These results show that the magnesium content of the populations of $\omega$\,Cen correlates with the abundances of oxygen and carbon and anti-correlates with helium and nitrogen. 
These correlations are interpreted as the result of proton-capture nucleosynthesis in the CNO cycle and the MgAl- chain of H-burning. 
 In particular, the MgAl- chain, which is responsible for magnesium variations is only active at temperatures higher than $\sim 7 \times 10^{7}$ K \citep[e.g.][]{denisenkov1990a, langer1993a, renzini2015a, prantzos2006a, prantzos2017a}.  
 Clearly, the material making the Mg-poor populations was exposed to higher temperatures than the Mg-rich stars, with p-captures having destroyed Mg making Al.
 
We find that about 70\% of the selected metal-intermediate stars of $\omega$\,Cen exhibit similar values of $\Delta_{\rm F275W,F280N}$ but are clustered around three different values $\Delta_{\rm F343N,F373N}$. We verified that these three groups of stars populate the lower, middle and upper streams defined by \citet{marino2019a} and are composed of stars with different  nitrogen contents.   
Accurate estimates of the magnesium abundance of metal-intermediate and metal-rich stars of $\omega$\,Cen are challenged by the internal variation of [Fe/H] in these stars. Nevertheless, the fact that a sample of metal-intermediate stars exhibit larger values of  $\Delta_{\rm F275W,F280N}$ than Mg-rich metal-poor stars indicates that they are depleted in [Mg/Fe].

\citet{norris1995a} determined abundances of magnesium and of other 19 elements for 40 RGB stars of $\omega$\,Cen, based on high-resolution spectroscopy. They find that the majority of stars have nearly constant [Mg/Fe]$\sim$0.4, with the exception of five stars where magnesium is clearly underabundant ([Mg/Fe]$\sim -0.1$). Two Mg-depleted stars have [Fe/H]$\sim -1.7$ while the other three have slightly higher iron abundances of [Fe/H]$\sim -1.5$. Hence, the group of Mg-depleted stars populate the metal-poor portion of the [Fe/H] distribution of the stars analyzed by \citet{norris1995a}, which span the interval $-1.8\lesssim$[Fe/H]$\lesssim-0.8$.
 \citet{norris1995a} noticed that a common property of the group is that it includes only CN-rich and Al-rich stars and that four out five stars are strongly oxygen depleted. 

Although {\it HST} photometry is not available for the stars studied by \citet{norris1995a}, their findings that the metal-poor population of $\omega$\,Cen hosts stars depleted by $\sim$0.5 dex in [Mg/Fe] with respect to the bulk of metal-poor stars and that the metal-intermediate stars also host stars with low magnesium abundances are consistent with the conclusions of our paper. 
The evidence that $\omega$ Cen hosts stars with depleted magnesium has been recently confirmed by \citet{meszaros2019a} based on spectra of 898 stars collected by the APOGEE survey. 
These results, based on high-resolution spectroscopy, demonstrate that the $\Delta_{\rm {\it C} F275W,F336W,F438W}$ vs.\,$\Delta_{\rm F275W,F814W}$ ChM introduced in this work, is an efficient tool to disentangle stellar populations with different [Mg/Fe] and infer their magnesium abundances.

\section*{acknowledgments} 
\small
We are grateful to the anonymous referee for a constructive report that has improved the quality of the manuscript.
This work has received funding from the European Research Council (ERC) under the European Union's Horizon 2020 research innovation programme (Grant Agreement ERC-StG 2016, No 716082 'GALFOR', PI: Milone, http://progetti.dfa.unipd.it/GALFOR), and the European Union's Horizon 2020 research and innovation programme under the Marie Sklodowska-Curie (Grant Agreement No 797100, beneficiary Marino). APM, MT and ED acknowledge support from MIUR through the FARE project R164RM93XW SEMPLICE (PI: Milone). APM and MT  have been supported by MIUR under PRIN program 2017Z2HSMF (PI: Bedin).
C.\,L.\. acknowledges support from the one-hundred-talent project of Sun Yat-set University. C.\,L.\, was supported by the National Natural Science Foundation of China under grants 11803048.
This work is based on observations made with the NASA/ESA {\it Hubble Space Telescope}, obtained from data archive at the Space Telescope Science Institute (STScI). STScI is operated by the Association of Universities for Research in Astronomy, Inc. under NASA contract NAS 5-26555.

\section*{Data availability}
The data underlying this article will be shared on reasonable request to the corresponding author.

\bibliography{ms}

\end{document}